\documentstyle[11pt,amssymb,epsf]{article}
\let\a=\alpha    
    
  \let\n=\nu

\let\C=\Chi

\def\nn{\nonumber} \def\bd{\begin{document}} \def\ed{\end{document}}
\def\ds{\documentstyle} \let\fr=\frac \let\bl=\bigl \let\br=\bigr
\let\Br=\Bigr \let\Bl=\Bigl 
\let\bm=\bibitem
\let\na=\nabla
\let\pa=\partial \let\ov=\overline 
\newcommand{\be}{\begin{equation}} 
\newcommand{\ee}{\end{equation}} 
\def\ba{\begin{array}}
\def\ea{\end{array}}
\def\ft#1#2{{\textstyle{{\scriptstyle #1}\over {\scriptstyle #2}}}}
\def\fft#1#2{{#1 \over #2}}
\def\del{\partial}
\def\vp{\varphi}
\def\sst#1{{\scriptscriptstyle #1}}
\def\oneone{\rlap 1\mkern4mu{\rm l}}
\def\td{\tilde}
\def\wtd{\widetilde}
\def\ie{\rm i.e.\ }
\def\dalemb#1#2{{\vbox{\hrule height .#2pt
        \hbox{\vrule width.#2pt height#1pt \kern#1pt
                \vrule width.#2pt}
        \hrule height.#2pt}}}
\def\square{\mathord{\dalemb{6.8}{7}\hbox{\hskip1pt}}}
\newcommand{\ho}[1]{$\, ^{#1}$}
\newcommand{\hoch}[1]{$\, ^{#1}$}
\newcommand{\bea}{\begin{eqnarray}} 
\newcommand{\eea}{\end{eqnarray}} 
\newcommand{\ra}{\rightarrow}
\newcommand{\lra}{\longrightarrow}
\newcommand{\Lra}{\Leftrightarrow}
\newcommand{\ap}{\alpha^\prime}
\newcommand{\bp}{\tilde \beta^\prime}
\newcommand{\tr}{{\rm tr} }
\newcommand{\Tr}{{\rm Tr} } 
\def\0{{\sst{(0)}}}
\def\1{{\sst{(1)}}}
\def\2{{\sst{(2)}}}
\def\3{{\sst{(3)}}}
\def\4{{\sst{(4)}}}
\def\5{{\sst{(5)}}}
\def\6{{\sst{(6)}}}
\def\7{{\sst{(7)}}}
\def\8{{\sst{(8)}}}
\def\n{{\sst{(n)}}}
\def\cA{{{\cal A}}}
\def\cF{{{\cal F}}}
\def\tV{\widetilde V}
\def\tW{\widetilde W}
\def\tH{\widetilde H}
\def\tE{\widetilde E}
\def\tF{\widetilde F}
\def\tA{\widetilde A}
\def\im{{{\rm i}}}
\def\tY{{{\wtd Y}}}
\def\ep{{\epsilon}}
\def\vep{{\varepsilon}}
\def\R{\rlap{\rm I}\mkern3mu{\rm R}}
\def\bD{{{\bar D}}}

\def\R{\rlap{\rm I}\mkern3mu{\rm R}}
\def\bD{{{\bar D}}}
\def\R{{{\Bbb R}}}
\def\C{{{\Bbb C}}}
\def\H{{{\Bbb H}}}
\def\CP{{{\Bbb C}{\Bbb P}}}
\def\RP{{{\Bbb R}{\Bbb P}}}
\def\Z{{{\Bbb Z}}}
\def\bA{{{\Bbb A}}}
\def\bB{{{\Bbb B}}}
\def\bC{{{\Bbb C}}}
\def\bZ{{{\Bbb Z}}}
\def\csch{{\rm csch}}
\def\cosec{{\,\hbox{cosec}\,}}

\newcommand{\tamphys}{\it Center for Theoretical Physics,
Texas A\&M University, College Station, TX 77843, USA}
\newcommand{\umich}{\it Michigan Center for Theoretical Physics,
University of Michigan\\ Ann Arbor, MI 48109, USA}
\newcommand{\upenn}{\it Department of Physics and Astronomy,
University of Pennsylvania\\ Philadelphia,  PA 19104, USA}
\newcommand{\SISSA}{\it  SISSA-ISAS and INFN, Sezione di Trieste\\
Via Beirut 2-4, I-34013, Trieste, Italy}
 
\newcommand{\ihp}{\it Institut Henri Poincar\'e\\
  11 rue Pierre et Marie Curie, F 75231 Paris Cedex 05}
 
\newcommand{\damtp}{\it DAMTP, Centre for Mathematical Sciences,
 Cambridge University\\ Wilberforce Road, Cambridge CB3 OWA, UK}
\newcommand{\itp}{\it Institute for Theoretical Physics, University of
California\\ Santa Barbara, CA 93106, USA}
 
\newcommand{\auth}{M. Cveti\v{c}\hoch{\dagger}, G.W. Gibbons\hoch{\sharp},
H. L\"u\hoch{\star} and C.N. Pope\hoch{\ddagger}}
 
\begin{document}
\begin{flushright}
\hfill{DAMTP-2001-25}\ \ \ {CTP TAMU-28/01}\ \ \ {UPR-955-T}\ \ \
{MCTP-01-39}\ \ \ {RUNHETC-2001-26}\\
{August 2001}\ \ \
{hep-th/0108245}
\end{flushright}

%\vspace{15pt}
 
\begin{center}
{ \large {\bf Cohomogeneity One Manifolds of Spin(7) and $G_2$ Holonomy}}

\vspace{5pt}
\auth
 
\vspace{3pt}
{\hoch{\dagger}\upenn}
 
\vspace{3pt}

{\hoch{\dagger}\it Department of Physics and Astronomy, Rutgers
University\\ 
Piscataway, NJ 08855, USA}

\vspace{3pt}
{\hoch{\sharp}\damtp}

\vspace{3pt}
{\hoch{\star}\umich}
 
\vspace{3pt}
{\hoch{\ddagger}\tamphys}

\vspace{3pt}

\underline{ABSTRACT}
\end{center}

   In this paper, we look for metrics of cohomogeneity one in $D=8$
and $D=7$ dimensions with Spin(7) and $G_2$ holonomy respectively.  In
$D=8$, we first consider the case of principal orbits that are $S^7$,
viewed as an $S^3$ bundle over $S^4$ with triaxial squashing of the
$S^3$ fibres.  This gives a more general system of first-order
equations for Spin(7) holonomy than has been solved previously.  Using
numerical methods, we establish the existence of new non-singular
asymptotically locally conical (ALC) Spin(7) metrics on line bundles
over $\CP^3$, with a non-trivial parameter that characterises the
homogeneous squashing of $\CP^3$.  We then consider the case where the
principal orbits are the Aloff-Wallach spaces $N(k,\ell)=SU(3)/U(1)$,
where the integers $k$ and $\ell$ characterise the embedding of
$U(1)$.  We find new ALC and AC metrics of Spin(7) holonomy, as
solutions of the first-order equations that we obtained previously in
hep-th/0102185.  These include certain explicit ALC metrics for all
$N(k,\ell)$, and numerical and perturbative results for ALC families
with AC limits.  We then study $D=7$ metrics of $G_2$ holonomy, and
find new explicit examples, which, however, are singular, where the
principal orbits are the flag manifold $SU(3)/(U(1)\times U(1))$.  We
also obtain numerical results for new non-singular metrics with
principal orbits that are $S^3\times S^3$.  Additional topics include
a detailed and explicit discussion of the Einstein metrics on
$N(k,\ell)$, and an explicit parameterisation of $SU(3)$.

\pagebreak
\setcounter{page}{1}

\tableofcontents
\addtocontents{toc}{\protect\setcounter{tocdepth}{3}}
\vfill\eject

\section{Introduction}

    Metrics of special holonomy are of considerable interest both in
mathematics and in physics.  They are special cases of Ricci-flat
metrics, whose holonomy groups are strictly smaller than the $SO(D)$
holonomy of a generic $D$-dimensional metric.  The irreducible cases
include Ricci-flat K\"ahler metrics in dimension $D=2n$, with holonomy
$SU(n)$, and hyper-K\"ahler metrics in dimension $D=4n$, with holonomy
$Sp(n)$.  Two further irreducible cases arise, namely $G_2$ holonomy
in $D=7$, and Spin(7) holonomy in $D=8$.  It is to these latter cases,
known as metrics of exceptional holonomy, that this paper is devoted.

   Our focus in this paper will be on non-compact metrics of
cohomogeneity one, in dimensions $D=7$ and 8.  The first complete and
non-singular such examples were obtained \cite{brysal}, and first
appeared in the physics literature in \cite{gibpagpop}.  They
comprised three metrics of $G_2$ holonomy in $D=7$, and one of Spin(7)
holonomy in $D=8$, and all four of these metrics are asymptotically
conical (AC).  Specifically, the metrics in $D=7$ are asymptotic to
cones over $S^3\times S^3$, $\CP^3$, and the six-dimensional flag
manifold $SU(3)/(U(1)\times U(1))$, while the metric in $D=8$ is
asymptotic to a cone over $S^7$.  In all cases the base of the cone
carries an Einstein metric, albeit not the ``standard'' one.
Topologically, the three $D=7$ manifolds are the spin bundle of $S^3$,
and the bundles of self-dual 2-forms over $S^4$ and $\CP^2$
respectively.

    The topology of the $D=8$ manifold is the chiral spin bundle of
$S^4$.  The homogeneous $S^7$ of the principal orbits can be described
as a (round) $S^3$ bundle over $S^4$, with the sizes of the $S^3$
fibre and the $S^4$ base being functions of the radial variable.  The
specific forms of these functions ensure that the metric is complete
on the chiral spin bundle of $S^4$, with the radius of the $S^3$
fibres approaching zero at short distance in such a way that one
obtains the required non-singular $\R^4$ bundle over $S^4$.

   Recently, further complete and non-singular non-compact 8-metrics
of Spin(7) holonomy were found \cite{cglpspin7}.  By contrast to the
example in \cite{brysal,gibpagpop}, the new metrics are asymptotically
locally conical (ALC), approaching the product of a circle and an AC
7-manifold locally at large distance.  The 7-manifold is a cone over
$\CP^3$.  The new metrics were obtained by writing a metric ansatz
with a more general parameterisation of homogeneous metrics on the
$S^7$ principal orbits, in which the $S^3$ fibres over $S^4$ can
themselves be ``squashed,'' with the $S^3$ described as an $S^1$
bundle over $S^2$.  This now gives functions in the metric ansatz,
parameterising the sizes of the $S^4$, $S^2$ and $S^1$.  First-order
equations for these functions were derived in \cite{cglpspin7}, which
can be viewed as the necessary conditions for Spin(7) holonomy, and
then the general solution was obtained.  Besides the previous AC
example of \cite{brysal,gibpagpop}, which of course is contained as a
special case, all the new solutions are ALC.  The ALC nature of the
large-distance behaviour arises because the function parameterising
the size of the $S^1$ tends to a constant at infinity.  The general
solution of the first-order equations has a family of non-singular
metrics, with a non-trivial continuous parameter (\ie a parameter over
and above the trivial scale size).  In general the manifold is again
the chiral spin bundle of $S^4$.  For general values of the parameter
the solution is quite complicated, and is expressed in terms of
hypergeometric functions; the manifolds were denoted by $\bB^+$ and
$\bB^-$ in \cite{cglpspin7}.  For a particular value of the parameter
the solution becomes much simpler, and is expressible in terms of
rational functions; this case was denoted by $\bB_8$ in
\cite{cglpspin7}.  One further complete solution arises; an isolated
example which is topologically $\R^8$, and denoted by $\bA_8$ in
\cite{cglpspin7}.

   In section 2, we consider a further generalisation of the ansatz
for Spin(7) metrics with $S^7$ principal orbits, in which the $S^3$
fibres over the $S^4$ base have ``triaxial'' homogeneous distortions,
implying that there will now be a total of four functions
parameterising the various radii.  We obtain first-order equations
that imply Spin(7) holonomy, and then we discuss the possible
solutions.  The previous examples in \cite{brysal,gibpagpop} and
\cite{cglpspin7} of course arise as special cases.  Although we have
not been able to obtain more general solutions analytically, we have
carried out an extensive numerical analysis of the equations.  We find
clear evidence for the existence of non-singular triaxial solutions,
in which there is a minimal $\CP^3$ surface (a bolt) at short
distance, with an ALC behaviour at infinity.  There is a non-trivial
one-parameter family of such regular solutions where the parameter can
be thought of as characterising the ``squashing'' of the minimal
$\CP^3$, viewed as an $S^2$ bundle over $S^4$.  If we denote the ratio
of the radius of $S^2$ over the radius of $\CP^2$ by $\lambda$, then
we find non-singular metrics for $\lambda^2\le 4$.  The special case
$\lambda^2=4$ corresponds to the ``round'' Fubini-Study metric on
$\CP^3$, and in this case the 8-metric is nothing but the complex line
bundle over $\CP^3$ contained in \cite{berber,pagpop1}, which has the
smaller holonomy $SU(4)$.  When $\lambda^2<4$, the new metrics exhibit
a behaviour reminiscent of the Atiyah-Hitchin \cite{atiyhitch}
hyper-K\"ahler 4-metric, with the three radial functions $a_i$ on
$S^3$ going from $a_1^2=0$, $a_2^2=a_3^2=$constant at the bolt, to
$a_3=$constant, $a_1^2\sim a_2^2\sim r^2$ at large radius $r$.  We
denote these new 8-manifolds of Spin(7) holonomy by $\bC_8$.

    In section 3, we examine 8-metrics of Spin(7) holonomy where the
principal orbits are the Aloff-Wallach homogeneous spaces $N(k,\ell)$,
which are $SU(3)/U(1)$ with the integers $k$ and $\ell$ specifying the
embedding of the $U(1)$ in $SU(3)$.  We begin, in section 3.1, by
reviewing some of the relevant properties of the Aloff-Wallach spaces
themselves.  In particular, we present a more explicit demonstration
than has previously appeared in the literature of the fact that for
generic $k$ and $\ell$, each $N(k,\ell)$ admits two inequivalent
Einstein metrics.  (The existence of an Einstein metric for each
$N(k,\ell)$ was proven in \cite{myw}; an explicit expression for one
such metric on $N(k,\ell)$ was given in \cite{casrom}, and the proof
that each generic $N(k,\ell)$ admits two Einstein metrics was given in
\cite{pagpop}.)  We also give an alternative proof of a result
following from \cite{casrom,pagpop} that every homogeneous Einstein
metric has weak $G_2$ holonomy.  In section 3.2, we give a discussion
of the global structures of the $N(k,\ell)$ spaces, focusing in
particular on the question of when a given such space admits a
description as an $S^3$ (as opposed to lens-space) bundle over
$\CP^2$.  This is important in what follows in section 4, where we
discuss details of 8-metrics of Spin(7) holonomy with $N(k,\ell)$
principal orbits.  The first-order equations for Spin(7) holonomy for
this class of metrics were obtained in \cite{cglphyper}.  In order to
have a non-singular such metric on an $\R^4$ bundle over $\CP^2$, it
is crucial that the collapsing fibres at short distance should be
$S^3$ and not a lens space.  We obtain an explicit analytical local
solution, which is ALC, for each choice of $N(k,\ell)$ principal
orbit.  We also give a discussion of numerical solutions, which
indicate the existence of complete examples with a non-trivial
parameter, and which include metrics that are asymptotically-conical
in a particular limit.

   In section \ref{g2section}, we turn to a consideration of more
general 7-metrics of $G_2$ holonomy.  We begin in section
\ref{g2cp2sec} by studying 7-metrics of $G_2$ holonomy on the $\R^3$
bundle of self-dual 2-forms over $\CP^2$.  These generalise the AC
example on this topology in \cite{brysal,gibpagpop}, whose principal
orbits are the flag manifold $SU(3)/(U(1)\times U(1))$, with two size
parameters as metric functions.  The more general ansatz that we
consider here has principal orbits of the same topology, but with
three, instead of two, sizes as metric functions.  Interestingly, the
first-order equations that follow from requiring $G_2$ holonomy turn
out to be the same as those that arise in four dimensions, for a set
of Bianchi IX hyper-K\"ahler metrics.  In that case the equations was
solved completely in \cite{begipapo}, and so we are able to use the
same procedure here.  As in the four-dimensional case, we find here
that the general solution gives irregular metrics, with regularity
attained only if two of the metric functions are equal, which reduces
the system to the already-known one in \cite{brysal,gibpagpop}.

   In section \ref{g2s4sec}, we briefly consider the possibility of
more general 7-metrics of $G_2$ holonomy where the principal orbits
are $\CP^3$.  Although we end up concluding that no possibilities of
greater generality than those considered previously in
\cite{brysal,gibpagpop} arise, we do nevertheless obtain as a
by-product a more elegant formulation of the already-known metrics.

   In section \ref{g2sixfunsec}, we study a general system of
equations for metrics of $G_2$ holonomy on the spin bundle of $S^3$.
Here, the principal orbits have the topology of $S^3\times S^3$.  A
rather general ansatz with six functions parameterising sizes in a
family of squashed $S^3\times S^3$ metrics was studied in
\cite{cglpg2,brgogugu}, where first-order equations implying $G_2$
holonomy were derived.  We find by means of a numerical analysis that
the only non-singular solutions occur when two pairs of metric
functions are set equal, leading to a truncation to a four-function
system that was discussed in \cite{brgogugu}, where an isolated
non-singular ALC solution was obtained explicitly.  Perturbative
arguments in \cite{brgogugu} suggested that a more general family of
non-singular ALC solutions, with a non-trivial parameter, should
exist.  These would be analogous to the 1-parameter family of ALC
Spin(7) metrics found in \cite{cglpspin7}.  Using a numerical
analysis, we also find evidence for the existence of such a
1-parameter family of non-singular solutions.  We denote these by
$\bB_7^+$ and $\bB_7^-$, with the isolated example found in
\cite{brgogugu} being denoted by $\bB_8$.

   After conclusions, we include a number of appendices.  Appendix A
contains a detailed discussion of the parameterisation of $SU(3)$ in
terms of generalised Euler angles, which is useful in our discussion
of the global structure of the Aloff-Wallach $N(k,\ell)$ spaces.

        Recent applications of Ricci-flat manifolds with special
holonomy in string and M-theory can be found in [14-45].

\section{New Spin(7) metrics with triaxial $S^3$ bundle over $S^4$}
 
    In \cite{cglpspin7}, new complete non-singular Spin(7) metrics on
the chiral spin bundle of $S^4$, and on $\R^8$, were constructed.
These metrics have cohomogeneity one, with principal orbits that are
$S^7$, with a transitively-acting $SO(5)\times U(1)$ isometry, and
they are asymptotically locally conical (ALC).  They were obtained by
generalising the original ansatz used in the AC example of
\cite{brysal,gibpagpop}, by describing $S^7$ as an $S^3$ bundle over
$S^4$, with radial functions in the metric parameterising the size of
the $S^4$ base, and the sizes of $S^2$ and the $U(1)$ fibres in a
description of $S^3$ as the Hopf bundle over $S^2$.  First-order
equations coming from a superpotential were then constructed, and the
general solution was obtained.  A 1-parameter family of non-singular
solutions on the chiral spin bundle over $S^4$ was obtained; these
were denoted by $\bB_8^+$, $\bB_8^-$ and $\bB_8$ in \cite{cglpspin7}.
It should be emphasised that the parameter in these solutions is
non-trivial, and not merely a scale size.  The general solution also
includes an isolated non-singular Spin(7) metric on $\R^8$; this was
denoted by $\bA_8$ in \cite{cglpspin7}.  The local form of the metric
in this example is in fact the same as the metric on $\bB_8$ in the
1-parameter family on the chiral spin bundle of $S^4$, but with the
range of the radial coordinate chosen differently.

   In this section, we shall generalise the construction in
\cite{cglpspin7}, by introducing a fourth radial function in the
cohomogeneity one metrics, so that the principal orbits are now $S^7$
described as a bundle of triaxially-squashed 3-spheres over $S^4$.
After calculating the curvature, we find that the potential in a
Lagrangian description of the Ricci-flat conditions can be derived
from a superpotential, and hence we obtain a system of first-order
equations for the four metric functions.  These are equivalent to the
integrability conditions for Spin(7) holonomy.  In fact the equations
that we obtain have also been found recently by Hitchin \cite{hitch},
using a rather different method.

\subsection{Ansatz and first-order equations}

   We begin by introducing left-invariant 1-forms $L_{AB}$ for the
group manifold $SO(5)$.  These satisfy $L_{AB}=-L_{BA}$, and
%%%%%
\be
dL_{AB} = L_{AC}\wedge L_{CB}\,.
\ee
%%%%%
The 7-sphere is then given by the coset $SO(5)/SU(2)_L$, where we take
the obvious $SO(4)$ subgroup of $SO(5)$, and write it (locally) as
$SU(2)_L\times SU(2)_R$.

    If we take the indices $A$ and $B$ in $L_{AB}$ to range over the
values $0\le A\le 4$, and split them as $A=(a,4)$, with $0\le a\le 3$,
then the $SO(4)$ subgroup is given by $L_{ab}$.  This is decomposed as
$SU(2)_L\times SU(2)_R$, with the two sets of $SU(2)$ 1-forms given by
the self-dual and anti-self-dual combinations:
%%%%%
\be
R_i = \ft12(L_{0i} + \ft12\ep_{ijk}\, L_{jk})\,,\qquad
L_i = \ft12(L_{0i} - \ft12\ep_{ijk}\, L_{jk})\,,
\ee
%%%%%
where $1\le i\le 3$.  Thus the seven 1-forms in the $S^7$ coset will be
%%%%%
\be
P_a \equiv L_{a4}\,,\qquad R_1\,,\qquad R_2\,,\qquad R_3\,.
\ee
%%%%%
It is straightforward to establish that
%%%%%
\bea
dP_0 &=& (R_1+ L_1)\wedge P_1 + (R_2+L_2)\wedge P_2 + (R_3+L_3)\wedge 
      P_3\,,\nn\\
dP_1 &=& -(R_1+ L_1)\wedge P_0 - (R_2-L_2)\wedge P_3 + (R_3-L_3)\wedge 
      P_2\,,\nn\\
dP_2 &=& (R_1- L_1)\wedge P_3 - (R_2+L_2)\wedge P_0 - (R_3-L_3)\wedge 
      P_1\,,\nn\\
dP_3 &=& -(R_1- L_1)\wedge P_2 + (R_2-L_2)\wedge P_1 - (R_3+L_3)\wedge 
      P_0\,,\nn\\
dR_1 &=& -2 R_2\wedge R_3 - \ft12 (P_0\wedge P_1 + P_2\wedge
P_3)\,,\nn\\
dR_2 &=& -2 R_3\wedge R_1 - \ft12 (P_0\wedge P_2 + P_3\wedge P_1)\,,\nn\\
dR_3 &=& -2 R_1\wedge R_2 - \ft12 (P_0\wedge P_3 + P_1\wedge P_2)\,.
\label{so5d}
\eea
%%%%%

   We are now in a position to write an ansatz for 
the more general metrics of Spin(7) holonomy on
the $\R^4$ bundle over $S^4$,
%%%%%
\be
ds_8^2 = dt^2 + a_i^2\, R_i^2 + b^2\, P_a^2\,.\label{8metans1}
\ee
%%%%%
From this, we find after mechanical calculations using (\ref{so5d})
that the conditions for Ricci-flatness can be derived from the
Lagrangian $L=T-V$, together with the constraint $T+V=0$, where
%%%%%
\bea
T&=& 2\a_1'\, \a_2' + 2\a_2'\, \a_3' + 2\a_1'\, \a_3' + 
   8(\a_1'+\a_2'+\a_3')\, \a_4' + 12{\a_4'}^2\,,\nn\\
V&=& \ft14 a_1^2\, a_2^2\, a_3^2\, b^4\, (a_1^2+a_2^2+a_3^2)
  +2b^8\, (a_1^4+a_2^4+a_3^4 -2 a_1^2\, a_2^2 -2 a_2^2\, a_3^2 -2
  a_1^2\, a_3^2)\nn\\
&& -12 a_1^2\, a_2^2\, a_3^2\, b^6\,,
\eea
%%%%% 
where $a_i=e^{\a_i}$, $b=e^{\a_4}$, and a prime denotes a derivative
with respect to $\eta$, defined by $dt=a_1^2\, a_2^2\, a_3^2\, b^8\,
d\eta$.  Reading off the DeWitt metric $g_{ij}$ from the kinetic
energy $T=\ft12 g_{ij}\, {\a^i}'\, {\a^j}'$, we find that the
potential $V$ can be written in terms of a superpotential $W$, as
$V=-\ft12 g^{ij}\, (\del W/\del\a^i)\, (\del W/\del\a^j)$, where 
%%%%%
\be
W= a_1\, a_2\, a_3\, (a_1+a_2+a_3)\, b^2 - 2 b^4\, (a_1^4 + a_2^4 +
a_3^4 - 2 a_1\, a_2 -2 a_2\, a_3 - 2 a_3\, a_1)\,.
\ee
%%%%%
This leads to the first-order equations ${\a^i}'= g^{ij}\, \del
W/\del\a^j$, which gives
%%%%%
\bea
\dot a_1 &=& \fft{a_1^2 - (a_2-a_3)^2}{a_2\, a_3} -
\fft{a_1^2}{2b^2}\,,\nn\\
\dot a_2 &=& \fft{a_2^2 - (a_3-a_1)^2}{a_3\, a_1} -
\fft{a_2^2}{2b^2}\,,\nn\\
\dot a_3 &=& \fft{a_3^2 - (a_1-a_2)^2}{a_1\, a_2} -
\fft{a_3^2}{2b^2}\,,\nn\\
\dot b &=& \fft{a_1+a_2+a_3}{4b}\,,\label{spin7fo}
\eea
%%%%%
where a dot denotes a derivative with respect to the original radial
variable $t$ appearing in the ansatz (\ref{8metans1}).  It is
straightforward to see that these are in fact the integrability
conditions for Spin(7) holonomy.\footnote{These equations were also
obtained recently by N. Hitchin, using a rather different construction
\cite{hitch}.}

\subsection{Some properties of the equations}

\subsubsection{Truncations to simpler systems}

   First, note that if we drop the terms associated with $b$, we get
precisely the first-order system that arises for triaxial Bianchi IX
metrics in $D=4$ \cite{gibpop} that admits the Atiyah-Hitchin metric
\cite{atiyhitch} as a solution.  This corresponds to a limit in which
the radius of the $S^4$ goes to infinity, so that we effectively
recover the equations for Atiyah-Hitchin times flat $\R^4$.  Some
properties of the Atiyah-Hitchin solutions are reviewed in Appendix
\ref{ahsumsec}.

  If instead we set an two of the $a_i$ equal, say $a_2=a_3$, and make
the redefinitions $a_2=a_3\longrightarrow 2a$, $a_1\longrightarrow 2b$,
$b\longrightarrow c$, we get precisely the first-order system of our
previous paper \cite{cglpspin7} on the new Spin(7) manifolds $\bA_8$,
$\bB_8$ and $\bB_8^\pm$, namely
%%%%%
\be
\dot a = 1 -\fft{b}{2a} - \fft{a^2}{c^2}\,,\qquad
\dot b = \fft{b^2}{2a^2} - \fft{b^2}{c^2}\,,\qquad
\dot c = \fft{a}{c} + \fft{b}{2c}\,.\label{simpfirstorder}
\ee
%%%%%
This system was solved completely in \cite{cglpspin7}.

   A third specialisation is to set $a_2=-a_3$.  It can be seen from 
(\ref{spin7fo}) that this will be consistent provided that we also 
impose $a_2=2b$.  We then have the metric ansatz 
%%%%%
\be
ds_8^2 = dt^2 + a_1^2\, R_1^2 + a_2^2\, (R_2^2 + R_3^2  + \ft14 P_a^2)\,,
\label{kahler}
\ee
%%%%%
and the truncated first-order equations are
the equations
%%%%%
\be
\dot a_1 = 4 -\fft{3 a_1^2}{a_2^2}\,,\qquad \dot a_2 =
\fft{a_1}{a_2}\,.
\ee
%%%%%
It is straightforward to solve these, to give
%%%%%
\be
ds_8^2 = \Big(1-\fft{\ell^8}{r^8}\Big)^{-1}\, dr^2 + r^2\,  
\Big(1-\fft{\ell^8}{r^8}\Big)\, R_1^2 + r^2 \, (R_2^2 + R_3^2 + \ft14
P_a^2)\,.\label{linbun}
\ee
%%%%
This is in fact precisely the Ricci-flat K\"ahler metric with 
$SU(4)\equiv$ Spin(6)
holonomy, on an $\R^2$ bundle over $\CP^3$.  The complete metric is
asymptotic to the cone over $S^7/\bZ_4$. The 6-metric with $a_2^2$
prefactor is in fact precisely the Fubini-Study metric on $\CP^3$.

\subsubsection{Some observations about the Spin(7) system}

  Let us define $w_i$ variables as in the Atiyah-Hitchin case
\cite{atiyhitch}, namely 
%%%%%
\be
w_1 = a_2\, a_3\,,\qquad w_2=a_3\, a_1\,,\qquad w_3=a_1\, a_2\,,
\label{wdef}
\ee
%%%%%
and a radial variable $\eta$ by $dt=a_1\, a_2\, a_3\, d\eta$ (exactly
as for Atiyah-Hitchin).  Also, define
%%%%%
\be
\beta \equiv b^2\,.
\ee
%%%%%
Then the first-order equations (\ref{spin7fo}) become
%%%%%
\bea
\fft{d(w_1+w_2)}{d\eta} &=& 4 w_1\, w_2 - \fft1{2\beta}\, [w_1\, w_2\,
(w_1 + w_2) + w_3\, (w_1^2+w_2^2)]\,,\nn\\
\fft{d(w_2+w_3)}{d\eta} &=& 4 w_2\, w_3 - \fft1{2\beta}\, [w_2\, w_3\,
(w_2+w_3)+ w_1\, (w_2^2+w_3^2)]\,,\nn\\
\fft{d(w_3+w_1)}{d\eta} &=& 4 w_3\, w_1 - \fft1{2\beta}\, [w_3\, w_1\,
(w_3+w_1)+ w_2\, (w_3^2+w_1^2)]\,,\nn\\
\fft{d\beta}{d\eta} &=& \ft12 (w_1\, w_2 + w_2\, w_3 + w_3\, w_1)\,.
\label{weqs2}
\eea
%%%%%

   This set of equations can be reduced to a single highly non-linear
second-order equation.  To do this, we first make the field
redefinitions
%%%%%
\be
X=w_3 - w_1\,,\qquad Y = w_2-w_1\,,\qquad
Z=w_2 + w_3\,.
\ee
We can then derive the following simple equations
%%%%
\be
\fft{d}{d\eta} \log(\fft{X}{Y}) = 4(Y-X)\,,\qquad
\fft{d}{d\eta} \log(X\, Y\, \beta^2) = 4Z\,,
\ee
%%%%
together with two more complicated equations for $\dot Z$ and
$\dot \beta$.  In terms of a new radial variable $\tau$, defined by
$d\tau=Z\, d\eta$, we therefore have
%%%%%
\be
\beta^2\, X\, Y\, Z^{-2} = c_0\, e^{4\tau}\,,
\ee
%%%%%
where $c_0$ is a constant of integration.  After the further
redefinitions
%%%%%
\be
U\equiv \fft{X-Y}{Z}\,,\qquad V\equiv \fft{X}{Y}\,,
\qquad \td\beta = \beta\, Z^{-1}\,,
\ee
%%%%%
and the introduction of another radial variable $z$ defined by
$dz= 8U\, V\,(V-1)^{-2}\, d\tau$, we find that $V$ is given by
%%%%%
\be
V= \fft{z+1}{z-1}\,,
\ee
%%%%%
and then the remaining two equations, for $U$ and $\td \beta$, become
%%%%%
\bea
U'&=&\fft{U^2-2z\, U +1}{2(z^2-1)} - \fft{(U^2-1)(z\,
U-1)}{16(z^2-1)\,\td\beta}\,,\nn\\
\td\beta'&=& \fft{\td \beta\, (U^2-1)}{2(z^2-1)\, U} -
\fft{z\, U^3 + U^2 + 3z\, U -5}{16(z^2-1)\, U}\,,
\eea
%%%%%%
where a prime means $d/dz$.  One can solve
algebraically for $\td\beta$ in the $U'$ equation, and substitute
it back into the $\td\beta'$ equation, thereby obtaining a
second-order non-linear equation for $U$:
%%%%%
\bea
&&2 (z^2-1)^2 \, \Big( U\, (U^2-1)\, (z\, U-1)\, U''  - 
(2z\, U^3 -3 U^2-4 z\, U +5)\, {U'}^2\Big)\nn\\
&&\qquad+ (z^2-1)\, \Big( 2(z\, U-1)^2\, (U^2+5) -(U^2-1)^2\Big)\, U' 
\nn\\
&&\qquad\qquad+ 2(z\, U-1)^2 \, (z\, U^3 -3U^2 + 3z\, U -1) =0\,.
\eea
%%%%%

   It is not clear how to solve this equation in general.  Here, we
just remark that two special solutions are $U=2\, (z+1)^{-1}$ and
$U=2(z-1)^{-1}$.  In fact both of these correspond to the
previously-known solution (\ref{linbun}) on the complex line bundle
over $\CP^3$.  For example, for $U=2(z+1)^{-1}$, after defining a
radial coordinate $r$ by
$r^8=\ft1{16}(z+3)^{2}\,(z+1)^{-1}$, one obtains (\ref{linbun}) with
$\ell^8=\ft12$.  (The $\bA_8$ and $\bB_8$ metrics found in
\cite{cglpspin7} are also special solutions of the more general
triaxial system we are studying here, but these all correspond to a
degeneration of the parameterisation in this section, with $Y=0$ or
$X=Y$.)

\subsection{Numerical analysis}

   Since we have not been able to solve the first-order equations
(\ref{spin7fo}) explicitly, we now turn to a numerical analysis of the
equations.  In the case of non-singular manifolds, the metrics are
defined on $\R^+\times G/H$, completed by the addition of a degenerate
orbit $G/K$ at short distance, where $K$ contains $H$.  The possible
cases are $G/K=SO(5)/U(2)=\CP^3$, $G/K=SO(5)/SO(4)=S^4$, or
$G/K=SO(5)/SO(5)=\oneone$, corresponding to a $\CP^3$ or $S^4$ bolt,
or a NUT, respectively.

    Our technique consists of first performing an analytic Taylor
expansion of the solution in the neighbourhood of the degenerate orbit
(\ie in the neighbourhood of the NUT or bolt at short distance).  When
making this expansion, we impose the necessary boundary conditions to
ensure that the metric can be regular there (in an appropriate
coordinate system).  At this stage we are left with a number of
undetermined coefficients in the Taylor expansions, and these
represent the free parameters in the general solution that is
non-singular near the NUT or bolt.  We then use these Taylor
expansions in order to determine initial conditions a short distance
away from the NUT or bolt, and then we evolve these data to large
distance in a numerical integration of the first-order equations
(\ref{spin7fo}).  In particular, we can study the evolution of the
solution as a function of the free parameters, and determine the
conditions under which the solution is non-singular.

   We find that the non-singular solutions are those where the metric
at large distance approaches either a cone over $S^7/Z_4$ (the AC line
bundle over $\CP^3$ case) or $S^7$ (the AC chiral spin bundle over
$S^4$), or else a circle splits off and approaches a constant radius,
while the other directions in the principal orbits grow linearly so
that the metric approaches $S^1$ times a cone over $\CP^3$ locally.
These are the ALC cases.  In fact, we find that generically the
non-singular solutions are ALC, with the AC behaviour arising only as
a limiting case.

   In cases where the initial conditions do not lead to an ALC or AC
structure at infinity, we find from the numerical analysis that a
singularity arises in the metric functions.  In other words, the set
of non-singular metrics corresponds to those cases where the choice of
initial conditions leads to an ALC or AC behaviour at infinity.

\subsubsection{Numerical analysis for $\CP^3$ bolts: New Spin(7)
metrics $\bC_8$}

    We can study the solution space for regular metrics in this case
as follows.  First, we seek a solution in the form of a Taylor series
in $t$, for small $t$, that exhibits the required short-distance
behaviour.  In the present context, where we are looking for solutions 
in which the
metric collapses to a $\CP^3$ bolt at $t=0$, we make an expansion
%%%%%
\be
a_i(t) = \sum_{n\ge 0} x_i(n)\, t^n\,,\qquad b(t)= \sum_{n\ge 0}
y(n)\, t^n\,,\label{taylor1}
\ee
%%%%%
where $x_1(0)=0$, implying that $a_1$ vanishes at $t=0$.  Thus $R_2$
and $R_3$ describe the directions on an $S^2$ bundle over the $S^4$
that is described by the $P_a$.  

    We find that the general Taylor expansion of this form has 2 free
parameters.  These can be taken to be $y(0)$, specifying the radius of
the $S^4$ base, and $x_2(0)$ (which is equal to $-x_3(0)$) specifying
the radius of the $S^2$ fibres, on the $\CP^3$ bolt at $t=0$.  One of
these two parameters is trivial, corresponding merely to a choice of
overall scale, and so without loss of generality we may take $y(0)=1$.
For convenience, we shall define $x_2(0)\equiv \lambda$, and so this
corresponds to a non-trivial adjustable parameter in the solutions
that are regular near the $\CP^3$ bolt.  Thus the metric restricted to
the bolt is
%%%%%
\be
ds_6^2 = \lambda^2\, (R_2^2 + R_3^2)  + P_a^2 \,.\label{cp3bolt}
\ee
%%%%%
Note that this family of homogeneous metrics on $\CP^3$ reduces to the
standard Fubini-Study metric if $\lambda^2=4$.  

    To the first few orders in $t$, we find that the Taylor-expanded
solution to (\ref{spin7fo}) is given by
%%%%%
\bea
a_1 &=& 4 t + \fft{(\lambda^4-40\lambda^2 -48)}{12\lambda^2}\, t^3
+\cdots\,,\nn\\
a_2 &=& \lambda + (1-\ft14\lambda^2)\, t + 
\fft{(3\lambda^4 -8\lambda^2 +48)}{32\lambda}\, t^2 + \cdots\,,\nn\\
a_3 &=& -\lambda + (1-\ft14\lambda^2)\, t - 
\fft{(3\lambda^4 -8\lambda^2 +48)}{32\lambda}\, t^2 + \cdots\,,\nn\\
b &=& 1 + \ft1{16}(12-\lambda^2)\, t^2 +\cdots\,.
\eea
%%%%%

    Using the Taylor expansions to set initial data at some small
positive value of $t$, we now evolve the equations (\ref{spin7fo})
forward to large $t$ numerically.\footnote{In order to set the initial
data accurately at a sufficient distance away from the singular point
of the equations at $t=0$, we typically perform the Taylor expansions
to tenth order in $t$.}  We find that the solution with the above
small-$t$ behaviour is regular provided that the non-trivial parameter
$\lambda$ is chosen to that
%%%%%
\be
\lambda^2\le 4\,.
\ee
%%%%%
The case $\lambda^2=4$ corresponds precisely to the situation we
arrived at in (\ref{kahler}).  Namely, setting $\lambda^2=4$ in the
$\CP^3$ bolt metric (\ref{cp3bolt}), we get precisely the Fubini-Study
metric on $\CP^3$.  In fact the solution when $\lambda^2=4$ is nothing
but the Ricci-flat K\"ahler metric given by (\ref{kahler}).  This is
the limiting AC case that we alluded to above.  It has an
``accidental'' decrease in its holonomy group from the Spin(7) of the
generic solution of (\ref{spin7fo}) to $SU(4)\equiv$ Spin(6), with a
consequence increase from 1 to 2 parallel spinors.

    When $\lambda^2 <4$, we get new non-singular solutions, which we
shall denote by $\bC_8$.  From the numerical analysis, we find that
now, the metric function $a_3$ tends to a constant value at large
distance, while all the others grow linearly.  Thus for $\lambda^2<4$
the solution is ALC.  The case where $\lambda^2$ becomes zero is a
Gromov-Hausdorff limit in which the metric becomes the product
$M_4\times \R^4$, where $M_4$ is the Atiyah-Hitchin metric.  The
$\lambda$ modulus space of the new solutions is depicted in Figure 1
below (we assume, without loss of generality, that $\lambda$ is
non-negative, so that regular solutions occur for $\lambda\le2$).

\begin{figure}[ht]
\leavevmode\centering
\epsfxsize=5in
\epsfbox{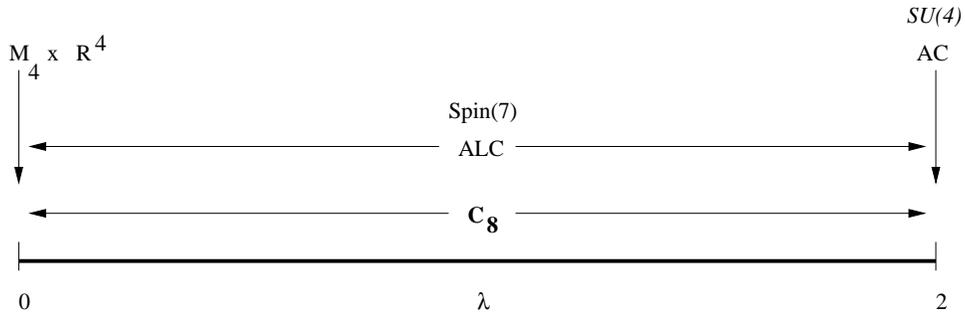}
\caption{The new non-singular Spin(7) metrics $\bC_8$ as a 
function of $\lambda$}
\end{figure}

    It should be noted that the new solutions exhibit the same
``slump'' phenomenon that was encountered in the Atiyah-Hitchin
metric.  Thus, at small distance it is the $a_1$ direction on $S^3$
that is singled out, with $a_1=0$ while $a_2=a_3=$constant on the
bolt.  By contrast, at large distance it is $a_3$ that is singled out
(by tending to a constant), while $a_1$ and $a_2$ become equal
asymptotically.  A sketch of the typical behaviour of the metric
functions $a_i$ and $b$ is given in Figure 2 below.

   It is worth remarking that the similarity of the asymptotic
behaviour of the new $\C_8$ metrics and the Atiyah-Hitchin metric may
have some interesting physical significance.  Like Atiyah-Hitchin, the
$\C_8$ metrics will have ``negative mass,'' as measured from infinity.
Just as the product of the Atiyah-Hitchin metric and seven-dimensional
Minkowski spacetime describes an orientifold plane in M-theory, so too
here we can expect that the $\C_8$ metrics will have an associated
interpretation in terms of orientifolds.

\begin{figure}[ht]
\leavevmode\centering
\epsfbox{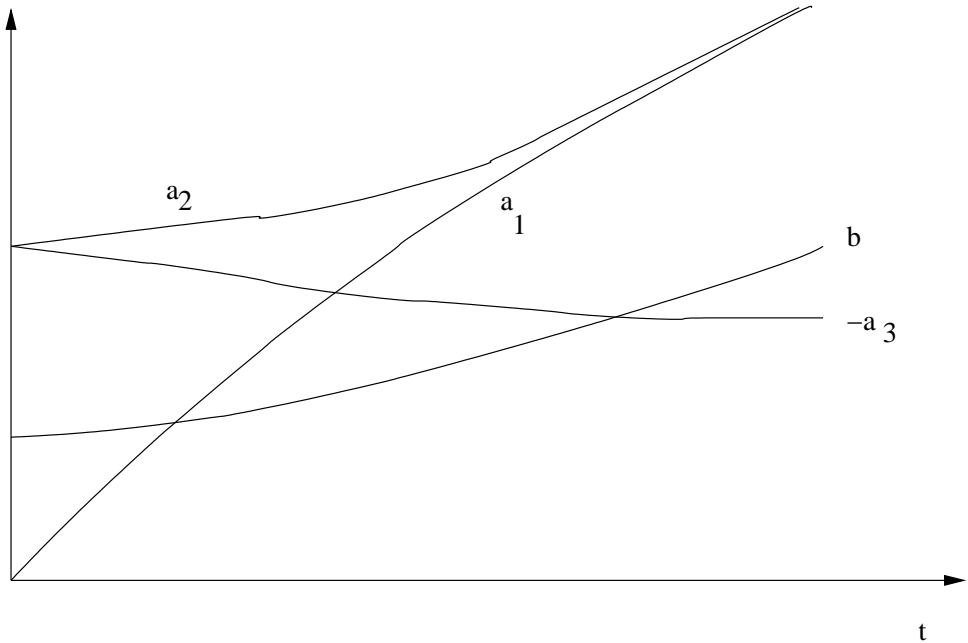}
\caption{The metric functions for a typical non-singular ALC solution}
\end{figure}

   The new non-singular Spin(7) metrics $\bC_8$ have the same topology
as the $\lambda=2$ example.  Thus they are line bundles over $\CP^3$
(specifically, the fourth power of the Hopf bundle).

\subsubsection{Numerical analysis for $S^4$ bolts: The $\bB_8$,
$\bB_8^+$ and $\bB_8^-$ examples recovered}

   This case can be studied by starting from the small-distance
expansion (\ref{taylor1}), and now taking $x_1(0)=x_2(0)=x_3(0)=0$.
We then find that the general such solution is characterised by three
parameters, which we shall relabel as $q_1$, $q_2$ and $q_3$.  The
first few orders give
%%%%%
\bea
a_1 &=& t - q_1\, t^2\ +\cdots\nn\,,\\
a_2 &=& t - q_2\, t^2\ +\cdots\nn\,,\\
a_3 &=& t - q_3\, t^2\ +\cdots\nn\,,\\
b &=& b_0 + \fft{3}{8 b_0}\, t^2 + \cdots\,,
\eea
%%%%%
where $b_0  = 2^{-1/2}\, \sqrt{q_1+q_2+q_3}$.  Note that we must have
$q_1+q_2+q_3>0$.  

    From the numerical solutions we find that regularity requires that
two of the $q_i$ be equal, leading, in turn, to the equality of the
corresponding pair of functions $a_i$.  Thus all the regular solutions
here reduce to ones that we have already found in \cite{cglpspin7}.
It is, nevertheless, of interest to see how they relate to the
previous results in \cite{cglpspin7}.

    Let us, without loss of generality, choose $q_2=q_3$.  This will
be understood to be the case in everything that follows.  The regular
solutions can then be summarised as follows.  In all cases we must
therefore have $q_1+2 q_2>0$.  Regularity also turns out to imply
$q_1>0$, and $q_2\le q_1$.  The cases are as follows:
%%%%%
\bea
-\ft12 q_1 < q_2 <0:&& \hbox{The $\bB_8^-$ metrics}\nn\,,\\
q_2=0:&& \hbox{The $\bB_8$ metric}\nn\,,\\
0< q_2 <q_1:&& \hbox{The $\bB_8^+$ metrics}\,,\nn\\
q_1=q_2:&& \hbox{The original AC Spin(7) metric of 
\cite{brysal,gibpagpop}}\,.
\eea
%%%%%
Note that when $q_2/q_1$ becomes equal to $-\ft12$, we have a
Gromov-Hausdorff limit to the product $M_7\times S^1$, where $M_7$ is
the original AC $G_2$ metric \cite{brysal,gibpagpop} 
on the $\R^3$ bundle over $S^4$.   The Spin(7) metrics are depicted in
Figure 3 below.

\begin{figure}[ht]
\leavevmode\centering
\epsfxsize=5in
\epsfbox{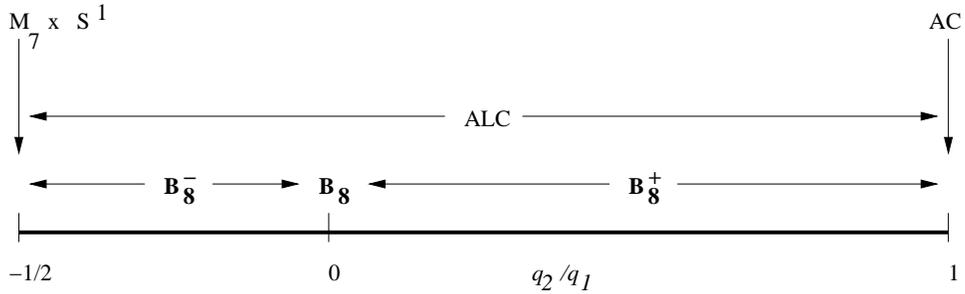}
\caption{The non-singular Spin(7) metrics $\bB_8$ and $\bB_8^\pm$ 
as a function of $q_2/q_1$}
\end{figure}

\subsubsection{Analysis for NUTs: The $\bA_8$ example recovered}

    The short-distance expansion for the NUT case corresponds to using
(\ref{taylor1}) with $x_1(0)=x_2(0)=x_3(0)=y(0)=0$.  On finds that
there are only two possible solutions at short distance.  One of these
is
%%%%%
\be
a_1=a_2=a_3=\ft35 t\,,\qquad b =\fft{3}{2\sqrt5}\, t\,,
\ee
%%%%%
which is an exact solution corresponding to the cone over the squashed
Einstein $S^7$.  It is singular at the apex.  The other solution has
the following expansion:
%%%%%
\bea
a_1 &=& -t + q\, t^3 -\ft32 q^2\, t^5 + \cdots\,,\nn\\
a_2 &=& a_3 = t\,,\nn\\
b &=& \ft12 t + \ft18 q\, t^3 -\ft{9}{64}\, q^2\, t^5+ \cdots\,.
\eea
%%%%%
The constant $q$ just corresponds to a trivial scale parameter here.
This is in fact precisely the previously known solution $\bA_8$.

    In summary, we get new regular Spin(7) metrics with squashed
$\CP^3$ bolts, with a non-trivial parameter $\lambda^2<4$
characterising the radius of the $S^2$ fibres relative to the $S^4$
base in the $\CP^3$ bolt.  If $\lambda^2=4$, we recover the
previously-known case of a complex line bundle over the Fubini-Study
metric on $\CP^3$, with $SU(4)$ holonomy, included in the cases
considered in \cite{berber,pagpop1}.  There are no new regular
solutions with $S^4$ bolts or for NUTS, since regularity in these
cases forces two of the $S^3$ directions to be equal, thus reducing
the systems to ones already solved in \cite{cglpspin7}.

\section{Spin(7) metrics with $SU(3)/U(1)$ principal orbits}

   In this section, we shall study the solutions of the first-order
equations for Spin(7) holonomy for metrics of cohomogeneity one whose
principal orbits are the Aloff-Wallach spaces $N(k,\ell)$, which are
cosets $SU(3)/U(1)$ where the integers $k$ and $\ell$ specify the
embedding of the $U(1)$.  All the necessary results for the
first-order Spin(7) equations were derived in \cite{cglphyper}, and
here we shall also follow the notation established in that paper.  We
begin our discussion with a study of the Aloff-Wallach spaces
themselves, and in particular the conditions under which they admit
Einstein metrics and metrics of weak $G_2$ holonomy.

\subsection{The principal orbits: Aloff-Wallach spaces}

   The coset spaces $SU(2)/U(1)$ are characterised by two integers,
$k$ and $\ell$, which specify the embedding of the $U(1)$ in $SU(3)$.
Specifically, if we represent $SU(3)$ by $3\times$ special unitary
matrices then the $U(1)$ subgroup can taken to be matrices of the form
%%%%%
\be
h=\pmatrix{ e^{\im k\, \theta}& 0 &0\cr
               0 & e^{\im\, \ell\, \theta} & 0\cr
                 0 & 0 & e^{-\im\, (k+\ell)\, \theta} }\,.\label{u1h}
\ee
%%%%%
The coset spaces are simply-connected when $k$ and $\ell$ are
relatively prime, and these are denoted by $N(k,\ell)$.  Clearly the
spaces $N(k,\ell)$, $N(\ell,k)$ and $N(k,-k-\ell)$ are topologically
identical, and in fact there is an $S_3$ permutation symmetry
generated by these two $Z_2$ operations.  It was shown in in
\cite{alowal} that all the $N(k,\ell)$ admit metrics of positive
sectional curvature.  Then, in \cite{myw}, an existence proof for an
Einstein metric on each $N(k,\ell)$ was given, and a more explicit
expression was found in \cite{casrom}.  Subsequently, it was shown in
\cite{pagpop} that each $N(k,\ell)$ in fact admits {\it two}
inequivalent Einstein metrics (except when $(k,\ell)=(0,1)$ or the
$S_3$-related values $(1,0)$ or $(1,-1)$, when there is only one).
Furthermore, it was proved that each such metric admits a Killing
spinor (except for one of the Einstein metrics with $k=\ell$, which
admits 3 Killing spinors).  The special case $k=\ell$ can be viewed as
an $SO(3)$ bundle over $\CP^2$, and the existence of the second
Einstein metric in this case had already been demonstrated in
\cite{jensen}.

\subsubsection{Einstein metrics on $N(k,\ell)$ from first-order 
equations}\label{d8fosec}

   Here, we present a summary of the construction of the Aloff-Wallach
spaces $N(k,\ell)$, and we give more explicit expressions for the
Einstein conditions than have been presented previously.  These will
be useful when we study cohomogeneity one Ricci-flat metrics with
$N(k,\ell)$ principal orbits in subsequent sections. A system of
first-order equations following from requiring Spin(7) holonomy for
such metrics was derived in \cite{cglphyper}, and we can first make
use of these in order to obtain equations for Einstein metrics on
$N(k,\ell)$.  For further details of the construction described below,
see \cite{cglphyper}.

   Defining left-invariant 1-forms $L_A{}^B$ for $SU(3)$, where
$A=1,2,3$, $L_A{}^A=0$, $(L_A{}^B)^\dagger =L_B{}^A$ and $dL_A{}^B =
\im\, L_A{}^C\wedge L_C{}^B$, we introduce the combinations
%%%%%
\bea
&&\sigma \equiv L_1{}^3\,,\qquad \Sigma\equiv L_2{}^3\,,\qquad \nu\equiv 
L_1{}^2\,,\nn\\
&&
\lambda\equiv  \sqrt2\, \cos\td\delta \, L_1{}^1 + \sqrt2\,
\sin\td\delta \, L_2{}^2\,,\label{ldefs}\\
&&Q\equiv -\sqrt2\, \sin\td\delta \, L_1{}^1 + \sqrt2\,
\cos\td\delta \, L_2{}^2\,,\nn
\eea
%%%%%
where $Q$ is taken to be the $U(1)$ generator lying outside the
$SU(3)/U(1)$ coset. It is evident by comparing with (\ref{u1h}) that
we have
%%%%%
\be
\fft{k}{\ell} = -\tan\td\delta\,,
\ee
%%%%%
and so $\td\delta$ is restricted to an infinite discrete set of values. 
Later, it will be convenient to write
%%%%%
\be
\cos\td\delta = -\fft{\ell}{\mu\, \sqrt2}\,,\qquad \sin\td\delta = 
\fft{k}{\mu\, \sqrt2}\,,\label{murel}
\ee
%%%%%
where $\sqrt2\, \mu\equiv \sqrt{k^2+\ell^2}$.

    In what follows we shall use real left-invariant 1-forms defined
by $\sigma=\sigma_1+\im\, \sigma_2$, $\Sigma=\Sigma_1 + \im\,
\Sigma_2$ and $\nu=\nu_1 +\im\, \nu_2$.  It was shown in
\cite{cglphyper} that if one defines 8-metrics of cohomogeneity one 
as follows:
%%%%%
\be
ds_8^2 = dt^2 + a^2\, \sigma_i^2 + b^2\, \Sigma_i^2 + c^2\, \nu_i^2 +
f^2\, \lambda^2\,,\label{d8ansatz}
\ee
%%%%%
where $a$, $b$, $c$ and $f$ are functions of the radial coordinate
$t$, then the first-order equations
%%%%%
\bea
\dot a &=& \fft{b^2+c^2-a^2}{b\, c} -
      \fft{\sqrt2\, f\, \cos\td\delta}{a}\,,\nn\\
\dot b &=& \fft{a^2+c^2-b^2}{c\, a} +
              \fft{\sqrt2\, f\, \sin\td\delta}{b}\,,\nn\\
\dot c &=& \fft{a^2+b^2-c^2}{a\, b} +
      \fft{\sqrt2\, f\,(\cos\td\delta -\sin\td\delta) }{c}\,,\nn\\
\dot f &=& - \fft{\sqrt2\, f^2\, (\cos\td\delta-\sin\td\delta)}{c^2}
   + \fft{\sqrt2\, f^2\, \cos\td\delta }{a^2} -
\fft{\sqrt2\, f^2\, \sin\td\delta}{b^2}\,,\label{fo2}
\eea
%%%%%
are the integrability conditions for Spin(7) holonomy.

   We can study Einstein 7-metrics on the principal orbits by taking
%%%%%
\be
a= \bar a\, t\,,\quad b=\bar b\, t\,,\quad c=\bar c\, t\,,\quad 
f=\bar f\, t\,,\label{conemet}
\ee
%%%%%
and solving the equations for the constants $\bar a$, $\bar b$,  $\bar
c$ and $\bar f$ that result from substituting (\ref{conemet}) into
the first-order equations (\ref{fo2}).   In other words, since we then
have
%%%%%
\be
ds_8^2 =dt^2 + t^2\, ds_7^2\,,
\ee
%%%%%
with
%%%%%
\be
ds_7^2 =  \bar a^2\, \sigma_i^2 + \bar b^2\, \Sigma_i^2 + 
\bar c^2\, \nu_i^2 + \bar f^2\, \lambda^2\,,\label{7metric}
\ee
%%%%%
it must be that if the 8-metric is Ricci-flat, then $ds_7^2$ is
Einstein, satisfying $R_{ab}=6\, g_{ab}$.  Furthermore, since the
first-order equations are the conditions for $ds_8^2$ to have Spin(7)
holonomy, it follows that $ds_7^2$ will have weak $G_2$ holonomy; in
other words it will not only be Einstein, but it will admit a Killing
spinor.  Since the results of \cite{casrom} showed that the Einstein
metrics discussed there admitted one or more Killing spinors, and the
results of \cite{pagpop} showed that all of the Einstein metrics on
the $N(k,\ell)$ spaces admit one or more Killing spinors, we will not
be losing any generality in our construction of Einstein metrics on
$N(k,\ell)$ by imposing the additional requirement of weak $G_2$
holonomy.  We shall, however, have the advantage of having a simpler
``first-order'' system of equations to work with.

    In order to simplify the notation, we shall drop the bar symbols
from the constants in the 7-metric (\ref{7metric}).  Thus after making
the substitution, we find that the metric
%%%%%
\be
ds_7^2 =  a^2\, \sigma_i^2 + b^2\, \Sigma_i^2 +  c^2\, \nu_i^2 +
f^2\, \lambda^2\,,\label{7metric2}
\ee
%%%%%
is Einstein, satisfying $R_{ab}=6 g_{ab}$, and of weak $G_2$ holonomy,
if the constants $a$, $b$, $c$ and $f$ satisfy the conditions
%%%%%
\bea
\fft{\ell\, f}{\mu\, a^2} &=& \fft{b^2+c^2-a^2}{a\, b\, c}-1\,,\nn\\
\fft{k\, f}{\mu\, b^2} &=& \fft{c^2+a^2-b^2}{a\, b\, c}-1 \,,\nn\\
 \fft{m\, f}{\mu\, c^2}&=& \fft{a^2+b^2-c^2}{a\, b\, c}-1\,,\nn\\
\Big(\fft{\ell}{a^2} + \fft{k}{b^2} + \fft{m}{c^2}\Big)
\, \fft{f}{\mu} &=& 1\,,\label{focons}
\eea
%%%%%
where in order to emphasise the symmetry, we have defined $m\equiv
-k-\ell$.  In fact, the system is invariant under the simultaneous
action of the permutation group $S_3$ on
$(\ell,k,m)$ and $(a,b,c)$.

   The permutation group $S_3$ can be generated by two $Z_2$ elements, 
namely 
%%%%%
\bea
A:&& k\longrightarrow \ell\,, \ \ \ell\longrightarrow k\,,\ \ 
   m\longrightarrow m\,,\nn\\ 
B:&& k\longrightarrow k\,, \ \ \ell\longrightarrow m\,,\ \ 
   m\longrightarrow \ell\,. 
\eea
%%%%%
If we define $x\equiv k/\ell$, then we shall have:
%%%%%
\bea
A:&& x\longrightarrow \fft{1}{x}\,,\nn\\
B:&& x\longrightarrow -\fft{x}{1+x}\,.\label{abz2}
\eea
%%%%%
It is easily seen that a ``fundamental domain'' 
%%%%%
\be
0\le x\le 1 \label{fundom}
\ee
%%%%%%
can therefore be chosen, with all other values of $x=k/\ell$
obtainable from this by acting with the $S_3$ permutation group.

    To solve the equations (\ref{focons}), we first note that two
independent relations involving only $a$, $b$ and $c$ can be derived,
one by adding all the equations, and the other by summing $a^2$ times
the first, $b^2$ times the second and $c^2$ times the third. Thus we
have
%%%%%
\be
a^2+b^2+c^2 = 4 a\, b\, c\,,\qquad a^4+b^4 +c^4 = 6 a^2\, b^2\, c^2\,.
\ee
%%%%%
It is straightforward to see that the general solution to these
equations can be written in terms of an angle $\phi$, such that
%%%%%
\be
a^2 = \fft{(2+\cos\phi)^2}{2(3+2\cos\phi+\sin\phi)}\,,\qquad
  b^2 = \fft{(2+\cos\phi)^2}{2(3+2\cos\phi-\sin\phi)}\,,\label{newsol0}
\ee
%%%%%
where
%%%%%
\be
0\le\phi < 2\pi\,.\label{phirange}
\ee
%%%%%
Substituting back into the remaining equations we therefore have
%%%%%
\bea
&&\fft{k}{\ell} = \fft{4+6\cos\phi + 12\sin\phi + 5\sin2\phi}{
                       4+6\cos\phi - 12\sin\phi - 5\sin2\phi}\,,
\qquad
c^2=\fft{(2+\cos\phi)^2}{8+ 12\cos\phi + 5\cos^2\phi}\,,\nn\\
&&f^2 = \fft{(2+\cos\phi)^2\, (\cos2\phi + 25 \cos^4\phi + 60
\cos^3\phi  -72\cos\phi -39)}{4(8 + 12\cos\phi + 5\cos^2\phi)^2}\,.
\label{newsol}
\eea
%%%%%
It should be recalled that we have normalised the Einstein metrics so
that they all have $R_{ab}=6 g_{ab}$. 

   Note that the set of solutions that we have obtained here maps into
itself under the action of the $S_3$ permutation group.  It is easily
seen that the $Z_2$ transformation $A$ in (\ref{abz2}) is implemented
by the replacement
%%%%%
\be
\phi\longrightarrow \phi' = 2\pi-\phi\,,
\ee
%%%%%
and this interchanges $a^2$ and $b^2$ (as well as $k$ and $\ell$),
while leaving $c^2$ fixed.  The $Z_2$ transformation $B$ in
(\ref{abz2}) is slightly trickier to implement.  It is achieved by
transforming from $\phi$ to $\phi'$ where
%%%%%
\be
\cos\phi' =
-\fft{2(1+\cos\phi-\sin\phi)}{3+2\cos\phi-\sin\phi}\,,\qquad
\sin\phi'= \fft{1+2\cos\phi + \sin\phi}{3+2\cos\phi-\sin\phi}\,.
\ee
%%%%%
This interchanges $a^2$ and $c^2$ (as well as implementing the mapping
on $k/\ell$ given in (\ref{abz2})), while leaving $b^2$ fixed.

   As $\phi$ varies over its range specified in (\ref{phirange}), the
function $k/\ell$ traverses each point on the real line exactly twice.
Of course the allowed values for $\phi$ are those for which the
expression for $k/\ell$ in (\ref{newsol}) are rational.  In general,
the two values $\phi_1$ and $\phi_2$ for $\phi$ that give the same
$k/\ell$ lead to {\it inequivalent} sets of values for the constants
$a$, $b$, $c$ and $f$, and hence to inequivalent Einstein metrics.
However, if it should happen that $\phi_1$ and $\phi_2$ are related by
$\phi_2=2\pi-\phi_1$, then it is evident from (\ref{newsol0}) and
(\ref{newsol}) that the associated pair of solutions will be
equivalent, with $a$ and $b$ interchanged.  This occurs only when
$k/\ell=-1$, and so in this case there is just one Einstein metric.
(The two values of $\phi$ that give rise to $k/\ell=-1$ are $\phi_1=
\arccos(-\ft23)$ and $\phi_2=2\pi-\arccos(-\ft23)$.)  Thus we have
reproduced the result of \cite{pagpop}, that each $N(k,\ell)$ space
has two inequivalent Einstein metrics, except for $N(1,-1)$, which has
only one.

   One final point remains.  We have defined the 1-form $\lambda$ as
in (\ref{ldefs}).  Despite naive appearances, this means that it is
not in fact normalised to a fixed length for arbitrary $\td\delta$.
This is because the ``metric'' for calculating the length is not
simply a $2\times2$ unit matrix in the $\C^2$ subspace spanned by
$L_1{}^1$ and $L_2{}^2$.  In fact, one should calculate lengths using
the $3\times3$ unit matrix in the $\C^3$ spanned by $L_1{}^1$,
$L_2{}^2$ and $L_3{}^3$, projected onto the plane defined by $L_1{}^1
+ L_2{}^2+L_3{}^3=0$ (the condition that ensures the $L_A{}^B$ are in
$SU(3)$ and not $U(3)$).

   The easiest way to calculate the length of $\lambda$ is therefore
to add the appropriate multiple of the $U(1)$ generator $U\equiv
L_1{}^1 + L_2{}^2 + L_3{}^3$ which lies in $U(3)$ but not in $SU(3)$,
such that the shifted 1-form $\td\lambda$ is orthogonal to $U$, which
implies:
%%%%%
\be
\td\lambda = \fft{\sqrt2}{3}\,\Big[ 
      (2\cos\td\delta-\sin\td\delta)\, L_1{}^1 +
      (2\sin\td\delta-\cos\td\delta)\, L_2{}^2 
     - (\cos\td\delta+\sin\td\delta)\, L_3{}^3\Big]\,.
\ee
%%%%%
Thus we see that the length of $\td\lambda$, and hence, by definition,
the length of $\lambda$, is given by
%%%%%
\be
|\lambda| = \ft{2}{\sqrt3}\, (1-\sin\td\delta\, \cos\td\delta)^{1/2}\,.
\label{lamlength}
\ee
%%%%%
Finally, since the $\lambda$ term appears in the metric via $ds_7^2= f^2\,
\lambda^2+\cdots$, and we now want to express this in terms of a
universally-normalised quantity
%%%%%
\be
\hat\lambda \equiv \fft{\lambda}{|\lambda|}\,,
\ee
%%%%%
so that $ds_7^2=\hat f^2\, \hat\lambda^2 + \cdots$, we see that we
should define
%%%%%
\be
\hat f^2 = \ft43(1-\sin\td\delta\, \cos\td\delta)\, f^2\,.
\ee
%%%%%
The quantity $\hat f$ will be invariant under the $S_3$ permutation group.
Written in terms of $k$, $\ell$ and $m=-k-\ell$, we have
%%%%%
\be
\hat f^2 = \ft43\, \fft{k^2+\ell^2+k\, \ell}{k^2+\ell^2}\, f^2 =
\ft23\, \fft{k^2+\ell^2+m^2}{k^2+\ell^2}\, f^2\,.\label{rescale}
\ee
%%%%%
The numerator factor $(k^2+\ell^2+m^2)$ is clearly invariant under the
permutation group, but the denominator $(k^2+\ell^2)$ is not.  It is
this non-symmetric denominator that corrects for the
non-permutation-invariance of $f$, making $\hat f$ permutation
invariant.  In fact we can see that if $f$ is replaced by $\hat f$ in
(\ref{focons}), then the factors of $\mu=\sqrt{k^2+\ell^2}/\sqrt2$ in
the denominators are precisely removed, so that the equations become
manifestly permutation symmetric, with $\hat f$ being invariant.

    It should be emphasised that using the Cartan-Maurer equation
$dL_A{}^B = \im\, A_A{}^C\wedge L_C{}^B$ we have $dU=0$, and so
therefore $d\lambda$ and $d\td\lambda$ are {\it identical}.  Thus
there is nothing wrong with our using $\lambda$ in our metric
constructions, it is just that its length is not given by the
expression one would naively have expected, but rather, by
(\ref{lamlength}).

\subsubsection{Einstein metrics on $N(k,\ell)$ from second-order
equations}\label{2einsec}

    Having constructed the Einstein metrics on $N(k,\ell)$ from the
first-order equations implying weak $G_2$ holonomy, it is instructive
now to re-examine the second-order Einstein equations themselves.  In
fact, as we shall show, these imply the previous first-order
equations, thus supplying another proof of the result in \cite{pagpop}
that all the Einstein metrics on $N(k,\ell)$ have weak $G_2$ holonomy,
and thus admit at least one Killing spinor.

   By following the same strategy as in the previous subsection, but
now calculating the conditions for Ricci-flatness of the cone over
$N(k,\ell)$, we find that the metric (\ref{7metric2}) $N(k,\ell)$ will
be Einstein, with Ricci tensor normalised to $R_{ab}=6g_{ab}$, if the
constants $a$, $b$, $c$ and $f$ satisfy
%%%%%
\bea
\fft{2 f^2\, \ell^2}{\mu^2\, a^4} &=& -6 + \fft{6}{a^2} + 
\fft{a^4-b^4-c^4}{a^2\, b^2\, c^2}\,,\nn\\
\fft{2 f^2\, k^2}{\mu^2\, b^4} &=& -6 + \fft{6}{b^2} + 
\fft{b^4-c^4-a^4}{a^2\, b^2\, c^2}\,,\nn\\
\fft{2 f^2\, m^2}{\mu^2\, c^4} &=& -6 + \fft{6}{c^2} + 
\fft{c^4-a^4-b^4}{a^2\, b^2\, c^2}\,,\nn\\
\fft{2f^2}{\mu^2}\, \Big(\fft{\ell^2}{a^4} + \fft{k^2}{b^4} +
\fft{m^2}{c^4}\Big) &=&6\,.\label{seceq}
\eea
%%%%%

  The approach to solving these equations that we shall present here
is an elaboration of the method that was presumably used in
\cite{casrom}.  Since a fully explicit derivation was not included
there, we shall give rather detailed results. We begin by introducing
new variables as in \cite{casrom}:
%%%%%
\be
A=\fft{a^2}{c^2}\,,\quad B=\fft{b^2}{c^2}\,,\quad u = \fft{\sqrt2\,
k\, f\, a}{\mu\, b\, c}\,,\quad
v=\fft{\sqrt2\, \ell\, f\, b}{\mu\, a\, c}\,,\quad
\lambda= \fft{6 a^2\, b^2}{c^2}\,.\label{casromvar}
\ee
%%%%%
In terms of these, the equations (\ref{seceq}) become
%%%%%
\bea
&&6A+B^2-A^2-1-u^2 = \lambda\,,\quad 6B+A^2-B^2-1-v^2 =
\lambda\,,\nn\\
&&6A\, B -A^2-B^2 -(A\, v+B\, u)^2 +1=\lambda\,,\quad
u^2+v^2+(A\, v+B\, u)^2 =\lambda\,.
\eea
%%%%% 
From these, we can obtain the following three equations in which
$\lambda$ is eliminated:
%%%%%
\bea
(2+B^2)\, u^2 + 2A\, B\, u\, v\, +(1+A^2)\, v^2 &=&
6A+B^2-A^2-1\,,\nn\\
(1+B^2)\, u^2 + 2A\, B\, u\, v + (2+A^2)\, v^2 &=&
6B+A^2-B^2-1\,,\nn\\
(1+2B^2)\, u^2 + 4A\, B\, u\, v +(1+2A^2)\, v^2 &=& 6A\, B -A^2-B^2
+1\,.
\eea
%%%%%
These can be viewed as three linear equations for the three quantities
$x\equiv u^2$, $y\equiv v^2$ and $z\equiv u\, v$.  After obtaining the
resulting expressions for $x$, $y$ and $z$ in terms of $A$ and $B$, we
then recall that $x\, y =z^2$, which leads to a polynomial 
constraint of the form $P(A,B)=0$.  In fact, we find that
%%%%%
\be
P(A,B) = \Big[ (A+B-3)^2 + 4(A-B)^2 -4\Big]\, Q(A,B)\,,\label{ppoly}
\ee
%%%%%
where $Q(A,B)$ is a sixth-order polynomial that can be shown not vanish for
any real positive $A$ and $B$ (see Appendix \ref{qsec} for the
expression for $Q(A,B)$, and
the proof of its positivity for nonvanishing $A$ and $B$).
Thus from $P(A,B)=0$ we conclude that
%%%%%
\be
 (A+B-3)^2 + 4(A-B)^2 =4\,,
\ee
%%%%%
which can be solved by writing $A+B-3=2\cos\phi$, $B-A=\sin\phi$, and hence
%%%%%
\be
A= \ft32 + \ft12\cos\phi -\sin\phi\,,\qquad B= \ft32 + \ft12\cos\phi +
\sin\phi\,.
\ee
%%%%%
The solutions for $u$ and $v$ then follow, giving
%%%%%
\be
u=\fft1{\sqrt2}\, (\cos\phi + 2\sin\phi)\,,\qquad v=\fft1{\sqrt2}
(\cos\phi -2\sin\phi) \,.
\ee
%%%%%
Finally, we find that
%%%%%
\be
\lambda = \ft32 (2+\cos\phi)^2\,,
\ee
%%%%%%
and hence from (\ref{casromvar}) we can obtain expressions for $a$,
$b$, $c$ and $f$.  These are in fact precisely the ones given in
(\ref{newsol0}) and (\ref{newsol}), which we previously obtained by
solving the conditions for weak $G_2$ holonomy.

   In summary, we have seen that the conditions (\ref{seceq})
following from imposing the Einstein equations have precisely the same
solution set as those coming from the simpler equations (\ref{focons})
that arose by requiring weak $G_2$ holonomy. 

\subsubsection{Global considerations}\label{globalsec}

   In order to investigate the global structure of the Spin(7) metrics
that we shall construct later, it is important to understand it first
in the Aloff-Wallach spaces themselves.  In particular, since in most
of our Spin(7) examples that we shall discuss below there will be a
$\CP^2$ degenerate orbit at short distance, it is important to
understand the structure of the Aloff-Wallach spaces {\it qua} bundles
over $\CP^2$.  Not surprisingly, triality plays an important role in
this question, and in fact a generic space $N(k,\ell)$ can be viewed
as any of three inequivalent such bundles.

    An example that is rather familiar is the case of the $N(1,1)$
space.  It is well known that $N(1,1)$ can be viewed as an $\RP^3$
bundle over $\CP^2$ (a physicist's discussion of this can be found in
\cite{pagpop}).  On the other hand, the principal orbits in the Calabi
metric on $T^*\CP^2$ are also the same Aloff-Wallach space, and so
clearly here it is being viewed as an $SU(2)$ bundle over $\CP^2$,
since the degeneration to the $\CP^2$ orbit in the Calabi metric is a
regular one, with the metric approaching $\R^4\times \CP^2$ locally.

   In fact in general it can be shown that if we view the $\sigma_i$
and $\Sigma_i$ 1-forms in (\ref{7metric2}) as spanning the $\CP^2$
base, and $\nu_i$ and $\lambda$ as spanning the 3-dimensional fibres,
then the space $N(k,\ell)$ can be described as an $S^3/Z_p$ lens-space
bundle over $\CP^2$, where
%%%%%
\be
p =|k+\ell|\,.\label{pkl}
\ee
%%%%%
(Note that $S^3/Z_0$ is a degenerate example, for which the fibres
will be $S^1\times S^2$.)   

    Applied to the case of $N(1,1)$ and its cousins $N(1,-2)$ and
$N(-2,1)$, we see that with respect to this convention choice of
having the $\nu_i$, together with $\lambda$, spanning the fibres, we
find that $N(1,1)$ will be an $S^3/Z_2=\RP^3$ bundle over $\CP^2$,
while $N(1,-2)$ and $N(-2,1)$ will be $S^3$ bundles over $\CP^2$.
This is consistent with the observations made above.

   There are various ways of proving the above results about the
topology of the bundles.\footnote{We are very grateful to James Sparks
and Nigel Hitchin for extensive help and discussions, and for
explaining how the result described above arises.}  Here, we shall
present a rather intuitive approach, base on a consideration of an
explicit parameterisation of $SU(3)$, which is presented in Appendix
A.

    We begin by recalling that $SU(3)/{\Bbb Z}_3$ acts effectively on
$\CP^2$, with stabiliser $U(2)$. Explicitly, if $(Z^1, Z^2, Z^3)$ are
homogeneous coordinates and $\zeta ^1= Z^1/Z^3 , \zeta ^2 = Z^2 /Z^3$
are inhomogeneous coordinates on $\CP^2$, then we may express almost
every element of $SU(2)$ as in appendix A, so that $(\phi, \theta ,
\psi, \xi)$ parameterise $\CP^2$, considered as the set of right
cosets, and $(\tilde \phi, \tilde \theta, \tilde \psi, \tau)$
parameterise the $U(2)$ stabiliser of the origin, $(0,0,1)$.  Note
that inhomogeneous coordinates $(\zeta ^1, \zeta ^2)$ are functions
just of $(\phi, \theta , \psi, \xi)$, and conversely. The $U(2)$
stabiliser of the origin is ${\tilde U}(\tilde \phi, \tilde \theta,
\tilde \psi) \, \exp (\im \sqrt 3 \tau \,\lambda _8)$ where the range
of the angles is $\tilde \phi \in (0,2 \pi]$, $\tilde theta \in [0,\pi
]$, $\tilde \psi \in (0, 4\pi]$ and $\tau \in (0, 2\pi]$. The two
coordinates $(\tau, \tilde \psi )$ label points in a maximal torus of
$U(2)$. A fundamental domain for the torus is given by a rectangle in
$\tau$-$\tilde \psi$ space of width $2\pi$ and height $4 \pi$.
 
    The circle $S^1_{k,\ell}$, parameterised by an angle $\a$, 
 may be expressed as
%%%%%
\be
\exp {\im \a \over 2} \{(k-\ell) \,\lambda_3 + (k+\ell) \sqrt 3
\lambda_8 \}\,,
\ee
%%%%%
and acting on the right it induces the action
%%%%%
\be
\tau \rightarrow \tau + (k-\ell) \, \a\,,\qquad
\tilde \psi \rightarrow  \tilde \psi + (k-\ell) \,\a\,. 
\ee
%%%%% 
If $k+\ell \ne 0$ (which will be treated separately), we may define a
coordinate $\Psi$ which is invariant under the circle action, and
which may be used to label its orbits, by  
%%%%%
\be
\tilde \psi \, \lambda_3 + \tau \,\sqrt 3 \,\lambda _8= { k-\ell \over
k+\ell}\, \tau \,\lambda _3 +
{ k+\ell \over k+\ell} \, \tau \, \sqrt 3 \, \lambda _8 +
\Psi \, \lambda _3\,.
\ee
%%%%%
One has
%%%%%
\be
\Psi = \tilde \psi  - { k-l \over k+l} \, \tau\,.
\ee
%%%%%
The problem is now to find the correct period for the angle
$\Psi$. This leads to a picture of $N(k,\ell)$ as a lens space bundle
over $\CP^2$.  The period is determined by the requirement that as
$\Psi$ ranges over its allowed values, it labels uniquely every orbit
of $S^1_{k,\ell}$ in $SU(3)$. To see how this is done it is helpful to
consider some examples.

   Let us consider $N(2,-1)$, for which we shall have $\Psi=\tilde
\psi - 3 \tau$. By examining the torus of side $2 \pi \times 4 \pi $
in $(\tau, \tilde \psi)$ space and following the orbit passing through
$(0,0)$ and its neighbours, it is easy to see that every orbit passes
once and only once through a strip of width $4 \pi \over 3$ in $\tau$
bounded by the straight lines $\Psi =0$ and $\Psi =-4 \pi $.  (The
verification of these facts is greatly assisted by drawing a diagram.)

   As another example, consider $N(1,0)$, for which we shall have
$\Psi=\tilde \psi -\tau$. Each orbit on the torus passes once and only
once through the square subdomain $ 0\le \tau \le 2 \pi$, $ 0 \le
\tilde \psi \2p$. The square lies inside the region bounded by
straight line $\Psi=2\pi $ and $\Psi= -2\pi $. Thus again the
range of $\Psi$ is $4 \pi$.

   As a third example, consider $N(3,2)$, which will give $\Psi=
\tilde \psi- { \tau \over 5} $. Following the orbit through the origin
around the torus we see that since it winds around the $\tau $
direction 10 times for every winding around the $\tilde \psi$
direction.  Thus the fundamental domain decomposes into ten strips of
height $2 \pi \over 5$ in $\tilde \psi$, and every orbit visits each
such strip once and only once. The range of $\Psi$ is therefore $2 \pi
\over 5$.  By applying similar arguments, one can fairly easily see
that in general, for $N(k,\ell)$, the period of $\Psi$ will be
$4\pi/|k+\ell|$.

   It is also of interest to identify what the bundles are.  For
example, we can think of $N(2,-1)$ or $N (1, -2) $ as the bundle of
unit cotangent vectors of $\CP^2$, \ie $ST^ \star\CP^2$. To see that
$SU(3)$ acts transitively, we need only remark that the stabiliser
$U(2) \subset SO(4) $ of a point in $\CP^2$ acts on the unit sphere in
${\Bbb R}^4$. We thus need to identify the stabiliser. In fact
$SU(2)\subset U(2) \subset SO(4)$ acts simply-transitively, and the
stabiliser is the circle action generated by an overall phase. In
terms of inhomogeneous coordinates, the action of $S^1_{k,\ell}$ is
$(\zeta ^1, \zeta ^2) \rightarrow (\exp {\im \theta (2k+\ell)}\,
\zeta^1, \exp {\im \theta \, (2\ell + k)}\, \zeta ^2 )$. It follows
that the cotangent vector at the origin of the form $(d \zeta ^1,0 )$
will be left invariant if and only if $2k+\ell=0$. In other words
$S^1_{1, -2}$ is the stabiliser of cotangent vectors at the origin
with vanishing second component. Similarly $S^1_{2, -1}$ is the
stabiliser of unit cotangent vectors at the origin with vanishing
first component. This is consistent with our result above that in the
case of $N(2,-1)$, the period of $\Psi$ is $4 \pi$.

   We have seen above that $N(2,-1)$ and $N(1,0)$ correspond to the
same bundle, since the period of $\Psi$ is the same.  This seems to be
related to the following: We can think of the cotangent bundle of
$\CP^2$ as the bundle of holomorphic 1-forms $\Lambda ^{1,0}$. Now
$\CP^2 $ has no Spin(4) structure, but it does have a Spin$^c$(4)
structure. One may identify the Spin$^c$(4) bundle with holomorphic
forms $\Lambda ^{\star,0}$. Under this identification, chirality
corresponds to Hodge duality. Thus the odd forms correspond to
negative chirality spinors.  It seems therefore that we may think of
both $N(2,-1)$ and $N(1,0)$ in terms of the the bundle of unit
negative-chirality spinors. The positive-chirality spinors correspond
to even forms, $\Lambda ^{0,0} \oplus \Lambda^{2,0}$. However, the
even forms are left invariant by the $SU(2)$ subgroup of the $U(2)$
stabiliser, and so even if normalised to have unit length they cannot
be a homogeneous space with respect to $SU(2)$. On the other hand, we
can consider the bundle of suitably charged negative-chirality
spinors. This amounts to giving the spinors a charge with respect to
the connection whose curvature is the K\"ahler form.

   To summarise, we see that in general $U(2)/S^1_{k,\ell}$ is a lens
space of the form $L(1,N) \equiv S^3/{\Bbb Z}_N$, where $N=|k+\ell |$.
For each geometrically distinct $N(k,\ell)$ space we will obtain in
general three different lens spaces, corresponding to the action of
the Weyl group $S_3$ of $SU(3)$. In particular cases we obtain fewer
than three bundles.  Thus for example $N(1,1)$ gives an $SO(3)\equiv
\RP^3$ bundle. On the other hand its Weyl cousins $N(-1,2)$ and
$N(2,-1)$ are both $SU(2) \equiv S^3$ bundles.

\subsection{An explicit Spin(7) solution for all 
$N(k,\ell)$}\label{explicit}

    Although we have not been able to obtain the general solution of
the first-order equations (\ref{fo2}) for Spin(7) metrics with
$N(k,\ell)$ principal orbits, we have succeeded in finding an isolated
exact solution to these equations for generic $k$ and $\ell$.  To
construct the solution, it is convenient to introduce a constant
$\gamma \equiv \tan\td\delta = -k/\ell$.  We find that there exists a
solution in which the following algebraic relation among the metric
functions $a$, $b$ and $c$ holds:
%%%%%
\be
X\equiv (1-2\gamma)\, a^2 + (\gamma-2)\, b^2 + (1+\gamma)\, c^2 + y=0\,,
\label{xdef}
\ee
%%%%%
where $y$ is a constant that sets the scale of the solution.  We shall
choose $y$ to be
%%%%
\be
y=8(\gamma-2)\,(\gamma+1)\,(2\gamma-1)=
\fft{8}{\ell^3}(k-\ell)(\ell-m)(m-k)\,,\label{ychoice}
\ee
%%%%%%% 
where as usual $m\equiv -k-\ell$.  By differentiating (\ref{xdef}) and
using the first-order equations, one obtains another algebraic
equation, which we may call $Y\equiv \dot X =0$.  This again involves
only $a$, $b$ and $c$, but not $f$.  Differentiating again, and using
the first-order equations, gives $W\equiv \dot Y=0$, which is an
algebraic equation for $a$, $b$, $c$ and $f$ (linear in $f$).  Thus
from these equations we can solve for $a$, $b$ and $f$ in terms of
$c$.  Differentiating again, we get $Z\equiv \dot W$ which again must
vanish for the solution.  However, this must be satisfied identically,
if the original supposition (\ref{xdef}) is correct, since otherwise
it would give us a solution for $c$ as a pure constant.  Calculation
shows that indeed $Z$ vanishes identically, so all is consistent, and
the validity of imposing the relation (\ref{xdef}) is established.

   It turns out now to be advantageous to work with a function $\rho$,
rather than $c$ itself, in order to avoid square roots, where $c^2 +
9(\gamma-1)^2=\rho^2$.  Thus the algebraic solutions for $a$, $b$, $c$
and $f$ in terms of $\rho$ are
%%%%%
\bea
&&a^2 = (\rho-\gamma-1)\,(\rho-\gamma+5)\,,\quad
b^2 = (\rho+\gamma+1)\,(\rho-5\gamma+1)\,,\nn\\
&&c^2=(\rho-3\gamma+3)\, (\rho+3\gamma-3)\,,\nn\\
&&
f^2 = \fft{9(\gamma^2+1)\, (\rho-5\gamma+1)\, 
      (\rho-\gamma+5)\, [\rho+3(\gamma-1)]}{
       2[\rho^2-(\gamma+1)^2]\, [\rho-3(\gamma-1)]}\,.\label{abcfgen}
\eea
%%%%%
We can now substitute these into any one of the first-order equations,
in order to obtain the differential equation for $\rho$.  (Since the
algebraic relations above were obtained by repeated use of the
first-order equations, there is only one remaining equation's worth of
information to be extracted from the entire first-order system, so we
can choose whichever of the four equations is most convenient.  The
$\dot f$ equation is a convenient choice.)  Using the coordinate gauge
choice $dt=f^{-1}\, dr$, we find that the differential equation for
$\rho$ is simply
%%%%%
\be
\rho' = \fft{\sqrt2}{3\sqrt{1+\gamma^2}}\,,
\ee
%%%%%
whose solution may be taken to be
%%%%%
\be
\rho = \fft{\sqrt2}{3\sqrt{1+\gamma^2}}\, r\,.
\ee
%%%%%
We see that $\rho$ is essentially just the radial coordinate, and the
metric can be written as 
%%%%%
\be
ds^2 = \ft92 (1+\gamma^2)\,
\fft{d\rho^2}{f^2} + a^2\, (\sigma_1^2+\sigma_2^2) +
b^2\, (\Sigma_1^2 + \Sigma_2^2) + c^2\, (\nu_1^2+\nu_2^2) +
f^2\, \lambda^2\,,\label{genmetrics}
\ee
%%%%%
where $a$, $b$, $c$ and $f$ are given by (\ref{abcfgen}).

   It is easy to see that the set of metrics we have obtained here
maps into itself under triality.  It is convenient to make use of this
observation when analysing the global properties; it allows us to
restrict attention to cases where the metric function $c$ is the first
of $a$, $b$ and $c$ to reach zero as $\rho$ reduces from the
asymptotic region at $\rho=\infty$.  The vanishing of $c$ will then
signal the inner endpoint of the radial coordinate range.  Before
studying this endpoint in detail, we may first observe that at large
distance the metrics are all asymptotically locally conical, since
$a$, $b$ and $c$ grow linearly, while $f$ tends to a constant.

   If $c$ is the first of $a$, $b$ and $c$ that vanishes as $\rho$
reduces from infinity, say at $\rho=\rho_0$, then it must be that the
factors in $a^2$ and $b^2$ in (\ref{abcfgen}) are still all positive
when $\rho$ reaches $\rho_0$ from above.  (We shall, without loss of
generality, assume that the asymptotic region is where
$\rho=+\infty$.)  It is easy to see from (\ref{abcfgen}) that for this
to happen we must have
%%%%%
\be
\rho_0= -3\gamma+3\,,\qquad \gamma\le -\ft12\,.
\ee
%%%%%
This in turn means that we must have $k/\ell\ge \ft12$, and since $k$
and $\ell$ must therefore have the same sign, we may without loss of
generality take them both non-negative.  We therefore have
%%%%%
\be
2k>\ell\ge 0\,.\label{klineq}
\ee
%%%%%
Noting from (\ref{pkl}) that the associated description of $N(k,\ell)$
will be as an $S^3/Z_p$ bundle over $\CP^2$ with $p=k+\ell$, it
follows that the case $p=1$ is achieved only if $\ell=0$, $k=1$.  The
conclusion from this is that the Spin(7) metrics (\ref{genmetrics})
will have $Z_p$ orbifold singularities on the $\CP^2$ bolt except in
the case that the principal orbits are $N(1,0)$.\footnote{The solution
in this special case $N(1,0)$ has been obtained by \cite{gukospar},
and we are very grateful to the authors for informing us of their
result prior to publication.  It has provided one of the motivations
for our investigations in this section.}

   If we consider a solution (\ref{abcfgen}) and (\ref{genmetrics})
for which $k$ and $\ell$ satisfy the inequality (\ref{klineq}), we
shall have an ALC Spin(7) metric with $N(k,\ell)$ principal orbits,
whose topology is an $R^4/Z_p$ bundle over $\CP^2$ with $p=k+\ell$.
Although this metric will be singular (if $p>1$), it is a relatively
``mild'' singularity, in the sense that it is an orbifold in which the
only infinities in the curvature will be delta-functions.  Such spaces
might in fact be relevant in string theory or M-theory, especially in
view of the fact that M-theory reductions could only give chiral
fermions if the reduction manifold is singular.

    Another possibility is that the orbifold singularities could be
resolved by considering more general metrics of higher cohomogeneity.
A phenomenon of this sort is known to occur in four dimensions, with
the multi-centre hyper-K\"ahler metrics.  The $N$-centre
multi-Eguchi-Hanson metric is complete and non-singular, and is
asymptotic to the cone over $S^3/Z_N$.  The $N=2$ example is nothing
but the Eguchi-Hanson metric, which can be written in its familiar
cohomogeneity one form.  However, the higher-$N$ metrics cannot have
cohomogeneity one.  Thus, for example, the metric one obtains by
imposing the periodicity $4\pi/N$ on the Hopf fibre coordinate in the
Eguchi-Hanson metric will have an orbifold singularity {\it qua}
metric of the Eguchi-Hanson cohomogeneity-one type, if $N>2$, but it
nevertheless admits a perfectly non-singular resolution as an
inhomogeneous multi-centre metric.  It may be that a similar situation
could arise with the resolution of metric on the cone over
$N(k,\ell)$.

\subsection{Small distance behaviour and numerical analysis}
\label{casessec}
 
    Since we have not been able to obtain the general solution to the
first-order equations (\ref{fo2}) analytically, we now turn to a
numerical analysis.  To begin we therefore need to construct Taylor
expansion for the solutions that are regular at small distance, \ie in
the region where one or more of the metric functions vanishes.  For
such an endpoint of the metric to be regular, it must be that the
terms that approach zero must be associated with the collapse of
spheres.  In the present case, we find that the possibilities that may
gave regular metrics are that $f$ alone vanishes, corresponding to a
collapse of circles, or else that $f$ and $a$ vanish, or $f$ and $b$
vanish, or $f$ and $c$ vanish, corresponding to a collapse of
3-spheres.  (These last three are equivalent, modulo the $S_3$
permutation group.)  One might also in principle have situations
with collapsing 2-spheres (just $a$ or just $b$ or just $c$
vanishing), or else with 5-spheres collapsing (by having $(a,b,f)$ or
$(a,c,f)$ or $(b,c,f)$ vanishing).
 
   We shall first describe the set of triality-related cases giving
collapsing 3-spheres (or lens spaces), where $(c,f)$ or $(a,f)$ or
$(b,f)$ vanish.  Thus the bolt at short distance will be $\CP^2$.  The
idea is to obtain Taylor expansions up to tenth order or so, which can
then be used in order to set initial data just outside the bolt, which
can then be integrated numerically to infinity.  We present just the
first couple of orders in the Taylor expansions here.

\subsubsection{A triality of short-distance solutions}\label{3casesec}

   Although we shall present the three possible classes of
small-distance solution, corresponding to $c$ or $a$ or $b$ vanishing
together with $f$, it should be emphasised that it is really redundant
to consider all three, since they are related by triality.  Thus one
can adopt two different viewpoints.  One possibility is to stick with
just one of the cases, say where $c$ and $f$ vanish, and then consider
all possible $N(k,\ell)$ principal orbits, including, in particular,
not only a given $N(k,\ell)$ but also its cousins $N(k,-k-\ell)$ and
$N(-k-\ell,\ell)$.  The other possibility is to consider all three
cases, with $(c,f)$, $(a,f)$ or $(b,f)$ vanishing, and then restrict
attention to a ``fundamental domain'' among the $N(k,\ell)$ spaces,
such as that defined in (\ref{fundom}).  Either viewpoint can be
taken, but one should take care not ``overcount'' the possibilities by
including all three cases and also including all the ``cousins.''  In
general, we shall find it convenient to adopt the first approach, and
consider all the $N(k,\ell)$ cousins within the framework of just Case
1 below.

\underline{\bf Case 1}:
\medskip

    First we consider the short-distance Taylor expansion
corresponding to the case where $c$ and $f$ vanish at $t=0$. We find
%%%%%%%
\bea
&&a=1 +\fft{5\cos\td \delta - 4\sin\td\delta}{6(\cos\td \delta -
          \sin\td\delta)} t^2 + \cdots\,,\qquad
b=1 + \fft{4\cos\td \delta - 5\sin\td\delta}{6(\cos\td \delta -
          \sin\td\delta)} t^2 + \cdots\,,\\
&&c=t + \ft{1}{\sqrt2}\Big(q (\cos\td\delta - \sin\td\delta)
-\ft{1}{\sqrt2}\Big)\, t^3  + \cdots\,,\qquad
f=-\fft{t}{\sqrt2(\cos\td\delta -\sin\td\delta)} + q\, t^3 +
\cdots\,.\label{case1}
\eea
%%%%%%

\underline{\bf Case 2}:
\medskip
 
   Now, we consider the triality-related case where it is instead $a$
that vanishes along with $f$ at $t=0$.  We find
%%%%%
\bea
&& a=t - (\ft1{\sqrt2} q\, \cos\td\delta +\ft12)\, t^3 +
\cdots \,,\qquad
b=1 + (\ft23 + \ft16\tan\td\delta)\, t^2 + \cdots\,,\nn\\
&& c=1 + (\ft56 -\ft16\tan\td\delta)\, t^2 +\cdots\,,\qquad
f=\fft{t}{\sqrt2\cos\td\delta} + q\, t^3 + \cdots\,.
\eea
%%%%%%%

\underline{\bf Case 3}:
\medskip

Finally, if $b$ instead vanishes along with $f$, we get
%%%%%%
\bea
&&a=1 + (\ft23 + \ft16 \cot\td\delta)\, t^2 + \cdots\,,\qquad
b=t + (\ft1{\sqrt2} q\, \sin\td\delta -\ft12)\, t^3\ + \cdots\,,\nn\\
&&c=1 + (\ft56-\ft16\cot\td\delta)\, t^2 + \cdots\,,\qquad
f=-\fft{t}{\sqrt2\sin\td\delta} + q\, t^3 + \cdots\,,
\eea
%%%%%%%
 
   It is easy to see that the three cases listed above are related by
triality.  

    Note that Case 1 is valid provided that $\cos\td\delta \ne
\sin\td\delta$, and likewise that Case 2 is valid for
$\cot\td\delta\ne 0$, and Case 3 for $\tan\td\delta\ne 0$.  These
exclusions are just a triality-related set.  Following our policy of
using just Case 1 for our discussion, we note that in this guise the
excluded case is $N(1,-1)$.  It is in fact easy to re-analyse the
Taylor expansion in Case 1 when $\cos\td\delta=\sin\td\delta$; we find
that we then get $f=0$ and the solution reduces to a Gromov-Hausdorff
limit of $S^1$ times a 7-metric of $G_2$ holonomy, whose principal
orbits are $S^2$ bundles over $\CP^2$.  ($G_2$ metrics of this type
will be discussed later, in section \ref{g2cp2sec}.)  This is
consistent with the general result discussed in section
\ref{globalsec}, where it was noted that the space $N(k,\ell)$ admits
a description as an $S^3/Z_p$ bundle over $\CP^2$, with $p=| k+\ell
|$.  Thus $N(1,-1)$ here corresponds to an $S^3/Z_0$ bundle over
$\CP^2$.  This bundle is a degenerate case, which is $S^1\times S^2$.

   Before discussing the numerical integration of (\ref{fo2}) using
these small-distance Taylor expansions to set up the initial data
outside the $\CP^2$ bolt, we first note that another situation of
particular interest is when the principal orbits are $N(1,1)$, or its
triality-related cousins $N(1,-2)$ or $N(-2,1)$.  Studying these
within the Case 1 framework, the example $N(1,1)$ (which is then
viewed as an $S^3/Z_2$ (\ie $\RP^3$) bundle over $\CP^2$, as can be
seen from (\ref{pkl})) arises as the principal orbits in the Spin(7)
manifold with $Z_2$ orbifold singularity that one gets by replacing
$S^4$ by $\CP^2$ in the chiral spin bundle over $S^4$ whose Spin(7)
metric was obtained in \cite{brysal,gibpagpop}.  Indeed, we find that
the Taylor expansion in (\ref{case1}) gives this exact solution if we
set $\tan\td\delta =-1$, and take the free parameter $q$ to have the
value $q=\ft16$, which implies $a=b$ and $f=-c/2$.
 
    The cousins $N(1,-2)$ and $N(-2,1)$ of $N(1,1)$ arise as the
principal orbits in the hyper-K\"ahler Calabi metric on $T^*\CP^2$.
As we shall discuss in more detail later, although this has the
smaller holonomy $Sp(2)=$ Spin(5), it is in fact a particular solution
of the Spin(7) first-order equations (\ref{fo2}), in the case of
$N(-2,1)$ and $N(1,-2)$.  These correspond respectively to
$\tan\td\delta = 2$ and $\tan\td\delta=\ft12$.  We find that the
Taylor expansions (\ref{fo2}) reduce to those for the exact Calabi
solution if the free parameter $q$ is chosen as follows:
%%%%%
\bea
\tan\td\delta = 2\,,\quad q = -\ft{\sqrt{10}}{3}:&&
a^2 + c^2 = b^2\,,\qquad a\, c = \sqrt{\ft25}\, b\, f\,,\nn\\
\tan\td\delta = \ft12\,,\quad q=\ft{\sqrt{10}}{3}:&&
b^2 + c^2 = a^2\,,\qquad 
b\, c = \sqrt{\ft25}\, a\, f\,.
\eea
%%%%%
 
\subsubsection{Results of numerical analysis}

    We are now in a position to make use of the series expansions of
section \ref{3casesec} to provide initial data just outside the
$\CP^2$ bolt, in order to perform a numerical integration of the
first-order equations (\ref{fo2}).  Again, because of the triality, we
need only discuss the series solution in Case 1, provided that we
consider $N(k,\ell)$ for all $k$ and $\ell$.  The discussion can be
further narrowed since the Case 1 is invariant under the $Z_2$
symmetry $k\leftrightarrow \ell$, $a\leftrightarrow b$,
$f\leftrightarrow -f$ and $q\leftrightarrow -q$.  It follows that we
need only concentrate on the cases with $|k|\le |\ell|$, implying that
we can consider the Case 1 solution with $|\tan\td \delta|\le 1$.

       The following is a summary of our numerical findings:

\noindent{\bf (a)} For each given $\tan\td\delta=-k/\ell$, there exist
a $q_0>0$, such that for parameters $q\ge q_0$, the functions
$(a,b,c,f)$ are regular.  In the limiting case where $q=q_0$, the
metric is AC, with the Einstein metric on $N(k\ell)$ on the base of
the cone being the one for which $\phi$, defined in (\ref{newsol}),
lies in the interval $[0, \pi)$.  For $q>q_0$, the metrics are ALC,
with $f$ becoming a finite constant at large distance.  The precise
value of $q_0$ for each $\td \delta$ is difficult to determine
numerically.  As we have seen previously, for $\tan\td\delta=-1$, we
have $q_0=\ft16$; and for $\tan\td\delta=\ft12$, we have
$q_0=\sqrt{10}/3$.
 
       Thus we see that for a generic value of $\td \delta$, there
exists an AC metric and a family of ALC metrics with a non-trivial
parameter, which are all regular aside from having an $\R^4/Z_p$
orbifold singularity on the $\CP^2$ bolt, where $p=|k+\ell|$.
In particular, this means that the AC metric and the ALC family of metrics
are completely non-singular in the case of $N(k,1-k)$ principal
orbits, for all integers $k$.
\medskip

\noindent{\bf (b)} Without loss of generality, we can restrict the
integers $k$ and $\ell$ so that $0<k\le l$, and then enumerate the
Case 1 solutions for all three cousins $N(k, \ell)$, $N(k, -k-\ell)$
and $N(\ell, -k-\ell)$.  For generic values of $k$ and $\ell$, the
three cousins will give rise to three distinctly different sets of AC
and ALC solutions.  If we focus in particular on the AC solutions,
then the choice $N(k, \ell)$ will be asymptotic to the cone over one
of the two Einstein metrics on $N(k,\ell)$, whilst its two cousins
$N(k, -k-\ell)$ and $N(\ell, -k-\ell)$ will give a pair of
(inequivalent) AC solutions that are asymptotic to the cone over the
{\it other} Einstein metric on this particular Aloff-Wallach space.
This implies that each given asymptotic cone structure admits two
different small-distance resolutions.

\underline{\bf Case 4}:
\medskip
 
    We have seen that the Case 1 solution is valid provided that
$\tan\td\delta\ne 1$.  When $\tan\td\delta=1$ two possibilities arise,
one of which is that $f=0$, as we discussed earlier.  Another
possibility is that $f$, as well as $a$ and $b$, becomes a constant at
small distance.  To the first few orders, we find that the solution is
given by
%%%%%
\bea
&&a=1-\ft13 q\, t + (1 -\ft5{18}q^2)\, t^2 +
(\ft7{45} -\ft{167}{810}q^2)\,q\, t^3 + \cdots\,,\nn\\
&&b=1+ \ft13 q\, t + (1-\ft5{18}q^2)\, t^2- 
(\ft7{45} - \ft{167}{810}q^2)\,q\, t^3 +\cdots\,,\nn\\
&&c=2t + \ft4{27}(q^2 -9)\, t^3 + \cdots\,,\qquad
f=q + \ft23 q^3\, t^2 + \cdots\,.\label{r3s5case}
\eea
%%%%%%%
(Owing to triality, there are also two additional (equivalent) types
of solution for either $\cos\td\delta=0$ or $\sin\td\delta=0$.)

    A numerical analysis shows that there exist regular solutions for
$|q|\le q_0=0.87\cdots$, such that the functions $a,b,c$ and $f$ are
regular as one integrates outwards.  When $|q|=q_0$, the solution is
AC, whilst for $|q|<q_0$, we have a non-trivial 1-parameter family of 
ALC solutions.  For $q=0$, we recover the case with $f=0$ mentioned above.

   In this class of solutions, where only the metric function $c$
vanishes at small distance, we see from (\ref{d8ansatz}) that we have
collapsing 2-spheres with metric described by $\nu_i^2$, whilst the
terms in $\sigma_i^2$, $\Sigma_i^2$ and $\lambda^2$ describe
homogeneous metrics on $S^5$ (viewed as an $S^1$ bundle over $\CP^2$).
Thus we see that at short distance the metrics approach an $\R^3$
bundle over $S^5$.  A straightforward calculation shows that the
squashed metric
%%%%%
\be
ds_5^2 = \sigma_i^2 + \Sigma_i^2 + x^2\, \lambda^2
\ee
%%%%%
on $S^5$ becomes the standard $SO(6)$-invariant round metric if
$x^2=1$.  From (\ref{r3s5case}) we see that our numerical result that
$|q|\le q_0=0.87\cdots$ therefore means that all the regular examples
arise when the $U(1)$ Hopf fibres on the $S^5$ bolt, viewed as an
$S^1$ bundle over $\CP^2$, are squashed relative to their length in
the round $S^5$ case.

\section{Analytic results for the Spin(7) equations for $N(k,\ell)$ orbits}

\subsection{The general case $N(k,\ell)$}

    It is advantageous first to rescale the metric function $f$, in
the fashion of (\ref{rescale}), so that the rescaled function is a
singlet under the $S_3$ permutation group.  Accordingly, we shall
define
%%%%%
\be
\td f \equiv \fft{\sqrt2}{\sqrt{k^2+\ell^2}}\, f\,.
\ee
%%%%%
Next, we define new variables $(A,B,F,G)$ in place of $(a,b,c,\td f)$
as follows:
%%%%%
\be
A=\fft{a^2}{c^2}\,,\qquad B=\fft{b^2}{c^2}\,,\qquad F=\fft{\td f\, a\,
b}{c^3}\,,\qquad G= \fft{a\, b}{c}\,.
\ee
%%%%%
These therefore satisfy the first-order equations
%%%%%
\bea
\dot A &=& \fft1{G}\, (4A - 4A^2 +2(k+\ell)\, A\, F +2\ell\, F)\,,\nn\\
\dot B &=& \fft1{G}\, (4B-4B^2 +2(k+\ell)\, B\, F + 2k\, F)\,,\nn\\
\dot F &=& \fft{F}{G}\, (5-3A-3B + 4(k+\ell)\, F)\,,\nn\\
\dot G &=& 3-A-B + (k+\ell)\, F +\fft{\ell\, F}{A} +\fft{k\, F}{B}\,.
\eea
%%%%%%

    If we now introduce a new radial variable $\eta$, related to $t$
by $dt=G\, d\eta$, these equations become
%%%%%
\bea
A' &=& 4A - 4A^2 +2(k+\ell)\, A\, F+2\ell\, F\,,\nn\\
B' &=& 4B-4B^2 +2(k+\ell)\, B\, F + 2k\, F\,,\nn\\
F' &=&(5-3A-3B)\, F + 4(k+\ell)\, F^2\,,\nn\\
G^{-1}\, G' &=& 3-A-B+(k+\ell)\, F 
    +\fft{\ell\, F}{A} +\fft{k\, F}{B}\,,\label{abfgeq}
\eea
%%%%%%
where a prime denotes a derivative with respect to $\eta$.  We see
that the first three equations now involve only the variables
$(A,B,F)$.  From these, one can solve for $F$ and $A$ in terms of $B$,
given by
%%%%
\bea
F&=&\fft{B' - 4B + 4B^2}{2(k + (k+\ell)\, B)}\,,\nn\\
A&=&\ft13 (k+ (k+\ell)\, B)^{-1}\, (B' - 4B +4B^2)^{-1}\, \Big(
-(k+ (k+\ell)\, B)\, B''\nn\\
&&(-11(2k+\ell)B + 9 (k+\ell)\,B^2 + 3(3k+(k+\ell)\, B'))\, B'\nn\\
&&+ 4 B\,(B-1) (5 k - 3 (2 k + \ell)\, B + 5 (k + \ell)\, B )\Big)\,.
\eea
%%%%
The system then reduces down to the following non-linear differential
equation for the function $B$:
%%%%%
\bea
&&18(k+\ell)\, B'^4 +3 (k + (k+\ell)B)^2\,
(7B''^2 - 3(-4B + 4 B^2 + B')\, B''')\nn\\
&& +
12 (2 k (4 k + 5 \ell) - (31 k^2  + 42 k\,\ell + 11\ell^2 ) B +
3(k+\ell)^2\,B^2)\, B'^3\nn\\
&&-64 B\,(-k^2  - k\, \ell\, B - \ell\, (k + \ell)\, B  +
(k + \ell)\,  B)\, B''\nn\\
&& +
4(k + (k+\ell)\, B)\, B')\,\Big((26 k - 2 \ell)\, B + 12 (k + \ell)\, B^2
 - 3 (6 k + (k + ll) B')\Big)\,B''\nn\\
&&8\Big(33k^2 -9k\, (17k+6\ell)\, B + (260k^2 + 260k\,\ell +77\ell^2)
B^2 \nn\\
&&-3(51k^2 + 92k\,\ell + 41\ell^2)\, B^3 + 57 (k+\ell)^2\, B^4\Big)\,
B'^2\nn\\
&&64B\, (B-1)\, \Big(12k^2 -7k\, (7k+2\ell)\,B +
3(22k^2 + 22k\,\ell + 7\ell^2)\,B^2 
\nn\\
&&- 7 (7k^2 + 12k\, \ell + 5\ell^2)\, B^3
+12(k+\ell)^2\, B^4\Big)\,B' + 128B^2\, (B-1)^2\, \Big(5k^2 - 3k
\, (7k+2\ell)\, B\nn\\
&&+(28k^2 + 28k\,\ell + 9\ell^2)\, B^2 -
3(7k^2 + 12k\,\ell + 5\ell^2)\, B^2 + 5(k+\ell)^2\, B^4\Big)
=0\,.
\eea
%%%%%%
Note that the equation is not explicitly dependent on the coordinate
$\eta$, implying that we can reduce the system to a third-order 
equation by defining $a\equiv B$ and $b(a)\equiv B'$.

\subsection{The case $N(1,-1)$}

\subsubsection{The first-order equations}

    Let us consider the Spin(7) equations specifically for the case of
$N(1,-1)$ principal orbits.  Setting $k=-\ell=1$, we have
%%%%%
\bea
&&\dot a = \fft{b^2+c^2-a^2}{b\, c} - \fft{f}{a}\,,\qquad
\dot b = \fft{c^2+a^2-b^2}{c\, a} + \fft{f}{b}\,,\nn\\
&&\dot c = \fft{a^2 + b^2-c^2}{a\, b} \,,\qquad
\dot f = \fft{f^2}{a^2} -\fft{f^2}{b^2}\,,
\eea
%%%%%
and (\ref{abfgeq}) then reduce to
%%%%%
\bea
&& A'  = 4A - 4A^2 -2F\,,\qquad 
 B'  = 4B-4B^2 + 2F\,,\nn\\
&& F' = F\, (5-3A-3B)\,,\qquad
G^{-1}\, G' = 3-A-B-\fft{F}{A} +\fft{F}{B}\,.
\label{4first}
\eea
%%%%%

    From these equations, one can derive the following third-order
equation for $B$:
%%%%%
\bea
&&128B^2\, (B^2-1)\, (9B^2-15B+5) + 88(7B^2-9B+3)\, {B'}^2 +
   8(14B-9)\, B'\, B'' \nn\\
&&+64B\, (B-1) \, (21B^2-35 B+12)\, B' -64B\, (B-1)\, B'' + 7{B''}^2 -
24{B'}^2 \nn\\
&& -3[B'+4B\, (B-1)]\, B'''=0\,.\label{thirdorder}
\eea
%%%%%

    If we now define a new variable $Q$ by $Q\equiv \sqrt{9-8B}$, 
we may note that the following give solutions of (\ref{thirdorder}):
%%%%%
\bea
\hbox{\bf (1)}:&&  Q'= \ft14(Q^2-1)(Q-3)\,,\nn\\
\hbox{\bf (2)}:&&  Q'= \ft14(Q^2-1)(Q+3)\,,\nn\\
\hbox{\bf (3)}:&& Q'= \ft14 Q^{-1}\, (Q^2-1)(Q^2-9)\,.
\eea
%%%%%
Here a prime means $d/d\eta$.  The solution (1) is
%%%%%
\be
B= \fft{1+2e^{2\eta} + \sqrt{1+e^{2\eta}}}{2(1+e^{2\eta})}\,,
\ee
%%%%%
and it is the special case for $N(1,0)$ of the explicit solutions found in
section \ref{explicit}, which was obtained in \cite{gukospar}.
The solution (2) corresponds to interchanging the roles of $A$
and $B$ relative to (1), and it has
%%%%%
\be
B= \fft{1+2e^{2\eta} - \sqrt{1+e^{2\eta}}}{2(1+e^{2\eta})}\,.
\ee
%%%%%
The solution (3) is rather trivial, and has 
%%%%%
\be
B= \fft1{1-e^{-4\eta}}\,,
\ee
%%%%%
which leads to $F=0$; it corresponds to a
degeneration of the metric to a 7-dimensional one.

   Another approach to solving the equations (\ref{4first}) is to
define
%%%%%
\be
A=X+Y\,,\quad B=X-Y\,.
\ee
%%%%%
The first three equations in (\ref{4first}) now give
%%%%%
\be
X'=4(X-X^2-Y^2)\,,\quad Y'=4Y-8X\, Y -2F\,,\quad F'=(5-6X)\, F\,.
\ee
%%%%%
Calculating $Y''$, using the other first-order equations, and then
using $d/d\eta=X'\, d/dX$, we get the second-order equation
%%%%%
\bea
&&(X-X^2-Y^2)\, \Big[4(X-X^2-Y^2)\, \fft{d^2Y}{dX^2} - 8Y\,
\Big(\fft{dY}{dX}\Big)^2 + (6X-5)\, \fft{dY}{dX}\Big] \nn\\
&&+ Y\, (5-8X+4X^2-8Y^2)=0\,.
\eea
%%%%%
Note that a special solution of this equation is 
%%%%%
\be
Y=\fft{\sqrt{1-X}}{\sqrt2}\,,
\ee
%%%%%
which gives rise to the explicit metric in section \ref{explicit} for
the case $N(1,0)$, which was obtained in \cite{gukospar}.

\subsubsection{Heuristic discussion of the flows}

   We can give the following analysis of the fixed-points of the
first-order equations (\ref{4first}).  Solving for $A'=B'=F'=0$, we
see that the fixed points occur for $(A,B,F)$ given by
%%%%%
\bea
&&(0,0,0)\,,\quad (0,1,0)\,,\quad (1,0,0)\,,\quad (1,1,0)\,,\nn\\
&& (\ft{5+\sqrt5}{6},\ft{5-\sqrt5}{6}, \ft{4\sqrt5}{9})\,,\quad 
(\ft{5-\sqrt5}{6},\ft{5+\sqrt5}{6}, -\ft{4\sqrt5}{9})\,.
\eea
%%%%%
It is easily seen that $(0,0,0)$ is a degenerate point, $(1,0,0)$ and
$(0,1,0)$ correspond to $\CP^2$ bolts at short distance, and $(1,1,0)$
is the large-distance asymptotic limit for ALC metrics.  The two
points $ (\ft{5+\sqrt5}{6},\ft{5-\sqrt5}{6}, \ft{4\sqrt5}{9})$ and
$(\ft{5-\sqrt5}{6},\ft{5+\sqrt5}{6}, -\ft{4\sqrt5}{9})$ correspond to
large-distance asymptotic limits for AC metrics.  The metrics on the
principal orbits in these last two limits are precisely the Einstein
metric on the $N(1,-1)$ Aloff-Wallach space, as discussed in section
5.1.

   The explicit ALC solution found in \cite{gukospar} (described in
section \ref{explicit} for $N(1,0)$) corresponds to a flow from
$(0,1,0)$ to $(1,1,0)$.  As we adjust the non-trivial constant which
parameterises inequivalent solutions of the first-order equations that
start from a $\CP^2$ bolt at short distance, we get a family of flows
that run from $(0,1,0)$ to the ALC endpoint at $(1,1,0)$.  As the
parameter is pushed to a limiting value, the distance at which the
function $f$ ``turns over'' and becomes asymptotically constant grows
larger and larger.  Eventually, at the limiting value of the
parameter, the distance at which this happens gets pushed to infinity,
and the endpoint jumps to the AC value
$(\ft{5+\sqrt5}{6},\ft{5-\sqrt5}{6}, \ft{4\sqrt5}{9})$.  If the
parameter is taken beyond the limiting value, the flow runs to some
singular point and the metric is correspondingly singular.

    The large-distance structure of the AC solution can be studied as
follows.  Let us suppose that we have the case
$(\ft{5-\sqrt5}{6},\ft{5+\sqrt5}{6}, -\ft{4\sqrt5}{9})$, in which $B$
approaches $\ft{5+\sqrt5}{6}$ asymptotically.  Setting
%%%%%
\be
B = \ft{5+\sqrt5}{6} + y(\eta)
\ee
%%%%%
in (\ref{thirdorder}), and then linearising in $y$, we obtain the
third-order equation
%%%%%
\be
9 y''' + 48 y'' -16 y' -160 y=0\,.\label{ylin}
\ee
%%%%%
Writing $y\sim e^{\lambda\, x}$, we find that the constant $\lambda$
must satisfy the auxiliary equation
%%%%%
\be
9\lambda^3 + 48\lambda^2 -16\lambda -160=0\,.\label{llin}
\ee
%%%%%
All three roots $\lambda_i$ are real, with $\lambda_1<0$,
$\lambda_2<0$ and $\lambda_3>0$.  Since we want solutions that
approach the AC limit (and hence $y\longrightarrow0$ as
$\eta\longrightarrow\infty$), we conclude that regular solutions must
have the asymptotic form
%%%%%
\be
y \sim x_1\, e^{\lambda_1\, \eta} + x_2\, e^{\lambda_2\, \eta}\,.
\ee
%%%%%

   We can think of the general solution as being characterised by 3
parameters (excluding the completely trivial constant shift of $\eta$,
but including the constant scaling).  We see that the solutions with
regular large-distance AC behaviour lie on a two-dimensional
submanifold of ingoing trajectories, parameterised by the constants
$x_1$ and $x_2$.  On the other hand, we know that at the bolt, the
solutions regular there also lie on a two-dimensional submanifold, of
outgoing trajectories.  Although we do not know analytically how to
interpolate between the two regions, we can argue on general grounds
that the intersection of the two-dimensional outgoing submanifold at
short distance, and the two-dimensional AC ingoing submanifold at
large distance, should occur along a curve.  (This family would really
be just a single non-trivial solution, since the single parameter
along the curve would be a ``trivial'' one.)  Thus we can expect a
solution that is regular on the bolt and also regular at an AC
infinity.  This same conclusion is also indicated by the numerical
solutions.

   We can, of course, repeat the above discussion for the general case
of $N(k,\ell)$ principal orbits.  The principles are the same as for
$N(1,-1)$ but the discussion is a little more involved since there are
now two non-trivial fixed points that describe the flows to cones over
the two inequivalent Einstein metrics on the Aloff-Wallach space.  We
find that at the linearised level, the analogue of (\ref{ylin}) is now
%%%%%
\be
y'''+ 4 y''\, (2+\cos\phi) - y'\, (2+\cos\phi)^2 - 4y\, (
15\cos^3\phi + 20\cos^2\phi + 12\cos\phi +8)=0\,,
\ee
%%%%%
where $\phi$ is the angle parameterising the Einstein metrics on the
Aloff-Wallach spaces, which was introduced in (\ref{newsol0}).
The solutions will therefore be of the form $y\sim e^{\lambda\, \eta}$ with
%%%%%
\be
\lambda^3 + 4 \lambda^2\, (2+\cos\phi) - \lambda\, (2+\cos\phi)^2 - 4 (
15\cos^3\phi + 20\cos^2\phi + 12\cos\phi +8)=0\,.\label{lamcub}
\ee
%%%%%
It is easy to see that this cubic polynomial in $\lambda$ has extrema
at two values of $\lambda$, one negative and the other positive, for
all values of $\phi$.  One can also see that the cubic is itself
respectively positive and negative at the two extrema.  This shows
that all three roots $\lambda_i$ of (\ref{lamcub}) are real, and that
one, $\lambda_1$, is certainly negative, and another, $\lambda_3$, is
certainly positive.  Together with the fact that the cubic is negative
at $\lambda=0$, for all $\phi$, we can deduce that the intervening root,
$\lambda_2$, is negative, and so for all $\phi$, two of the
$\lambda_i$ are negative and one is positive.  Thus again we have a
two-dimensional submanifold of ingoing trajectories, supporting the
indications from the numerical analysis that there will be regular AC
solutions.

\subsection{Perturbative construction of AC metrics}\label{resolvsec}

      We have obtained evidence by means of a numerical analysis that
for each choice of $N(k,\ell)$ principal orbit, there are two possible
AC Spin(7) metrics, which approach the cones over the two inequivalent
Einstein metrics on $N(k,\ell)$.  The only exception is $N(1,0)$, for
which there is only one AC solution, since here there is only one
possible Einstein metric on the base of the cone.

    We have already alluded to the fact that for the special case of
principal orbits that are $N(1,1)$, or its cousins $N(1,-2)$ and
$N(-2,1)$, we actually know of two explicit AC solutions of the
first-order equations (\ref{fo2}).  One such solution is the complete
and non-singular hyper-K\"ahler Calabi metric on $T^*\CP^2$, which
happens to have the smaller holonomy group $Sp(2)$, but nonetheless
corresponds to a solution also of the Spin(7) first-order equations
(\ref{fo2}).  If we make our usual choice where it is the function $c$
in (\ref{8metans1}), rather than $a$ or $b$, that vanishes at short
distance, then the principal orbits will be $N(1,-2)$ or $N(-2,1)$ in
this case. The other exact solution is the Spin(7) metric that one
obtains by replacing $S^4$ by $\CP^2$ in the original construction in
\cite{brysal,gibpagpop} of the complete and non-singular Spin(7) AC
metric on the chiral spin bundle of $S^4$.  After the replacement, the
metric will have a $Z_2$ orbifold singularity on the bolt, since now
we have $N(1,1)$ as principal orbits, which can be described as an
$S^3/Z_p$ bundle over $\CP^2$ with $p=|k+\ell|=2$, \ie an $SO(3)$
bundle over $\CP^2$.  Nonetheless, the associated solution of the
first-order equations is non-singular, and from a physical point of
view in string theory, one might even find the orbifold singularity
attractive.

   Leaving aside for now the question of the acceptability or
otherwise of an orbifold singularity, we can take the two exact
solutions described in the previous paragraph as starting points for
perturbative constructions of AC solutions of the first-order
equations (\ref{fo2}), for values of $k/\ell$ that are close to the
values occurring in the exact solutions.  Thus, for example, we can
take the $N(1,1)$ solution with the $Z_2$ orbifold singularity, and
then seek a solution with $k/\ell=1-\ep$, order by order in $\ep$.  Of
course we should ultimately have in mind that $\ep$ should be
rational, but this does not present any difficulty.

   The other starting point with an exact AC solution is the
hyper-K\"ahler Calabi metric.  In this particular instance we find it
more convenient, rather than following our usual strategy of working
with $N(1,-2)$ or $N(-2,1)$ principal orbits in the framework where
$c$ vanishes on the bolt, to work instead in the framework where $a$
vanishes on the bolt, in which case we again have $N(1,1)$ principal
orbits.  Thus the perturbative expansions around both of the exact
solutions can be parameterised by taking $\tan\td\delta = -1 +\ep$.

\bigskip\bigskip
\noindent\underline{Case (a): Expansion around $SO(3)$ bundle over $\CP^2$}:
\bigskip
 
    Here we take as our zeroth-order starting point the Spin(7) metric
on the chiral spin bundle of $S^4$, given in \cite{brysal,gibpagpop},
but with $S^4$ replaced by $\CP^2$.  The principal orbits are
$N(1,1)$, with the metric function $c^2$ that multiplies $\nu_i^2$ in
(\ref{8metans1}) vanishing on the bolt.  We shall work up to and
including order $\ep^2$ in the expansion around $k/\ell=1$.
 
   In order to simplify our results for the perturbative expansion it
is helpful to introduce a new radial variable $\rho$, defined in terms
of $r$ by $\rho=r^{2/3}$.  After some algebra, we find that the
perturbative expansion up to order $\ep^2$ is given by
%%%%%%
\bea
a &=& \ft3{\sqrt{10}}\, \rho^{3/2}\Big(1 +
\fft{3\rho^{11} -11\rho^6 + 33 \rho -25}{264\rho\, (\rho^5-1)^2}
\, (2\epsilon + \epsilon^2) + {\cal O}(\epsilon^3)
\Big)\,,\nn\\
b &=& \ft3{\sqrt{10}}\, \rho^{3/2}\Big(1 -
\fft{3\rho^{11} -11\rho^6 + 33 \rho -25}{264\rho\, (\rho^5-1)^2}
\, (2\epsilon + \epsilon^2) + {\cal O}(\epsilon^3)
\Big)\,,\nn\\
c &=& \ft35\rho^{-1}\, (\rho^5-1)^{1/2}\,\Big( 1+
c_2\, \epsilon^2 + {\cal O}(\epsilon^3)\Big)\,,\nn\\
f&=& -\ft{3}{10} \rho^{-1}\, (\rho^5-1)^{1/2}\,\Big(1+
f_2\, \epsilon^2 + {\cal O}(\epsilon^3)\Big)\,,\nn\\
h&=&\fft{\rho^{5/2}}{(\rho^5-1)^{1/2}}\, \Big(1 +
h_2\, \epsilon^2 + {\cal O}(\epsilon^3)\Big)\,,
\eea
%%%%%%%
where
%%%%
\bea
c_2 &=& \fft{(\rho-1)^4\, v_1}{2613600\rho^2\, (\rho^5-1)^5} +
\rho^{-1}\,(\rho^5-1)\, u + (\rho^5-1)^{-1}\, \td u\,,\nn\\
f_2 &=& \fft{(\rho-1)^4\, v_2}{2613600\rho^2\, (\rho^5-1)^5} -
\rho^{-1}\,(\rho^5-1)\, u + (\rho^5-1)^{-1}\, \td u\,,\nn\\
h_2 &=& \fft{(\rho-1)^4\, v_3}{2613600\rho^2\, (\rho^5-1)^5} -
\ft13\rho^{-1}\,(\rho^5-1)\, u - (\rho^5-1)^{-1}\, \td u\,,
\eea
%%%%%%
and the functions $u$ and $\td u$ are given by
%%%%%%
\bea
u &=& -\fft{7}{165 \rho}\, _2F_1[1,\ft15,\ft65,\rho^{-5}] +
   \fft{7}{660 \rho^4}\, _2F_1[1,\ft45,\ft95,\rho^{-5}]\,,\nn\\
\td u &=& k + \ft{7}{59400} \rho^3\, (15 \rho^5-24\rho^2 -40)\nn\\
&& + \fft{7(\rho^5-1)^2}{495}\, \Big[ \rho^{-2}\, 
_2F_1[1,\ft15,\ft65,\rho^{-5}] - \ft14 \rho^{-5}\, 
 _2F_1[1,\ft45,\ft95,\rho^{-5}]\Big]\nn\\
&&- \ft{35}{4356} \Big[ \log(\rho^5-1) + 10\rho^{-1}\, 
_2F_1[1,\ft15,\ft65,\rho^{-5}] -\ft52 \rho^{-2}\, 
_2F_1[1,\ft25,\ft75,\rho^{-5}]\Big]\,.
\eea
%%%%%%%
Note that we have
%%%%
\bea
\fft{\del u}{\del \rho} &=& \fft{7(\rho^3-1)}{165(\rho^5-1)}\,,\nn\\
\fft{\del\td u}{\del\rho} &=& \ft7{495}\rho^{-1} -
\fft{175\rho^2\, (\rho-1)^2}{4356(\rho^5-1)} +
\fft{(1+8\rho^5 - 9 \rho^{10})\, u}{3\rho^2}\,.
\eea
%%%%%%%
The functions $v_i$ are polynomials in $\rho$, given by
%%%%%
\bea
v_1&=& 31250 + 187500\rho + 873820\rho^2 + 2495280\rho^3 +
5456950\rho^4 + 9894075\rho^5\nn\\
&& + 15688150\rho^6 + 21497477\rho^7 + 25980358\rho^8 + 27795095\rho^9 + 
26221340\rho^{10}\nn\\
&& + 21387495\rho^{11} + 14948034\rho^{12} + 8557431\rho^{13} 
+ 3870160\rho^{14} +1498795\rho^{15}\nn\\
&&+ 1344660\rho^{16} + 2431196\rho^{17} + 3781844\rho^{18} + 
4420045\rho^{19} + 4281340\rho^{20}\nn\\
&& +3511270\rho^{21} +  2437848\rho^{22} + 1389087\rho^{23} + 
693000\rho^{24} + 277200\rho^{25} + 69300\rho^{26}\,,\nn\\
%%%%
v_2&=& -156250 - 502500\rho - 1192820\rho^2 - 2381280\rho^3 
- 4221950\rho^4 - 6069165\rho^5\nn\\
&& - 7561010\rho^6 - 7585363\rho^7 - 5030102\rho^8 + 1216895\rho^9
+  10072100\rho^{10}\nn\\
&&+ 21255735\rho^{11} + 32192394\rho^{12} + 40306671\rho^{13}
+ 43023160\rho^{14} + 40712455\rho^{15}\nn\\
&& + 33048900\rho^{16} + 22612556\rho^{17} + 
11983484\rho^{18} + 3741745\rho^{19} - 1745300\rho^{20}\nn\\
&& - 3900290\rho^{21} - 3838392\rho^{22} - 2674773\rho^{23} - 
1524600\rho^{24} - 609840\rho^{25} - 152460\rho^{26}\,,\nn\\
%%%%%
v_3 &=&  125000 + 252500\rho + 38110\rho^2 - 862560\rho^3 
- 2793900\rho^4 - 5655745\rho^5\nn\\
&& - 10651680\rho^6 - 16695647\rho^7 - 22701588\rho^8 - 
27583445\rho^9 - 29937510\rho^{10}\nn\\
&&- 28501325\rho^{11} - 24368354\rho^{12} - 18632061\rho^{13}
-12385910\rho^{14} - 7010765\rho^{15}\nn\\
&&- 3391240\rho^{16} - 1197516\rho^{17} - 99774\rho^{18} +
231805\rho^{19} + 256540\rho^{20}\nn\\
&& + 223750\rho^{21} + 160032\rho^{22} + 91983\rho^{23} + 
46200\rho^{24} + 18480\rho^{25} + 4620\rho^{26}\,.
\eea
%%%%%%%%%
Here $k$ in $\td u$ is an integration constant, which should be
chosen to be
%%%%
\be
k=\ft{343}{59400} + \ft{35}{4356}\, \gamma + \ft{35}{4356}
[2\psi(\ft15)-\psi(\ft25)] 
\ee
in order that the solution be regular at small distance, where
$\gamma$ is the Euler-Mascheroni constant, and $\psi(z)\equiv
\Gamma'(z)/\Gamma(z)$ is the digamma function.

      At large distance, the functions become
%%%%%%
\bea
a&=&\ft{3}{\sqrt{10}}\, r\, (1 + \ft1{44}\epsilon + \ft1{88}\epsilon^2)
\,,\qquad
b=\ft{3}{\sqrt{10}}\, r\, (1 - \ft1{44}\epsilon - \ft1{88}\epsilon^2)
\,,\nn\\
c&=&\ft{3}{5}\, r\, (1 + \ft{21}{3872}\,\epsilon^2)\,,\qquad
f=-\ft{3}{10}\, r\, (1 + \ft{489}{3872}\, \epsilon^2)\,,\nn\\
h&=&1 + \ft{31}{11616} \epsilon^2\,,
\eea
%%%%%%%

            As a verification, one can check that the above cone
metric matches precisely to the conifolds obtained in section 3,
up to $\epsilon^2$ order, expanding around $\phi=0$.   At small
distance,  it can be matched to the Case 1 in section \ref{casessec},
with the constant $q$ specified as
%%%%%
\be
q=\ft16 + \Big(\ft{2225}{26136} - \ft{7}{495}\,
\pi\, \sqrt{1 + \ft2{\sqrt5}}\,\Big)\,\epsilon^2 +
{\cal O}(\epsilon^3)
\,.
\ee
%%%%%%

    What we have seen emerging here is an orderly expansion of the
metric functions around their unperturbed form, with corrections at
order $\ep$ and $\ep^2$ that are perfectly regular both at short
distance and at large distance.  This provides further evidence, of an
analytical nature, for the existence of regular solutions of the
first-order equations (\ref{fo2}) for AC metrics (\ref{8metans1}) with
Spin(7) holonomy, where the principal orbits are $N(k,\ell)$ with
general values of $k$ and $\ell$.  Of course one should distinguish
between having regular solutions of the first-order equations, and
having regular metrics, since, as we know, our starting point for the
perturbation series in this case is a metric with a $\Z_2$ orbifold
singularity.  Thus our emphasis in this specific perturbation
expansion is really with the regularity of the metric functions,
rather than with the complete regularity of the 8-metrics.
Nonetheless, as we mentioned previously, even those with orbifold
singularities on the bolt may be of interest in string theory and
M-theory.  However, the main point emerging here is that we see strong
supporting evidence for the proposition that there exist regular AC
solutions of the first-order equations, for all $k$ and $\ell$.  For
those cases where $|k+\ell|=1$ (which are, of course, far away from
the $\ep=0$ starting point here), we should therefore obtain complete
and non-singular AC metrics.

\bigskip\bigskip
\noindent\underline{Case (b): Expansion around Calabi metric}:
\bigskip

    In this case we take as our zeroth-order starting point the
hyper-K\"ahler Calabi metric on $T^*\CP^2$, which is complete and
non-singular.  Since we shall choose to work in a framework where it
is the metric function $a$ that vanishes on the bolt, and since the
$S_3$ symmetry of the system involves permuting $(a,b,c)$ in step with
$(\ell,k,m)$, where $m=-k-\ell$, it follows that instead of
$N(k,\ell)$ being viewed as an $S^3/Z_p$ bundle over $\CP^2$ with
$p=|k+\ell|=|m|$ as it is when $c$ vanishes on the bolt, we now have
an $S^3/Z_p$ bundle with $p=|\ell|$.  Thus the non-singular
hyper-K\"ahler Calabi metric is described in these conventions in
terms of $N(1,1)$ or $N(-2,1)$ Aloff-Wallach spaces forming the
principal orbits.  We shall take $N(1,1)$, so that again our
perturbation will be of the form $\tan\td\delta=1-\ep$.

     It should be emphasised that although our starting-point here is
hyper-K\"ahler, we perturb around it using the usual Spin(7)
first-order equations.   Thus the reduced holonomy when $\ep=0$ is to
be viewed as an ``accidental'' reduction that is a feature of this
specific solution of the Spin(7) equations. 

   After some algebra, we find in this case that up to order $\ep^2$,
the perturbed solution is given by
%%%%%%
\bea
a&=&\sqrt{\ft12(r^2-1)}\, (1 + a_1\, \epsilon + 
a_2\, \epsilon^2)\,,\nn\\
b&=&\sqrt{\ft12(r^2+1)}\, (1 + b_1\, \epsilon + 
b_2\, \epsilon^2)\,,\nn\\
c&=& r\, (1 + c_1\, \epsilon + c_2\, \epsilon)\,,\nn\\
f&=&\ft12r \sqrt{1-r^{-4}}(1 + f_1\, \epsilon +
f_2\, \epsilon^2)\,,\nn\\
h&=&\fft{r}{2f}\,,
\eea
%%%%
where
%%%%
\bea
a_1&=& \fft{1+3z-3z^2}{12z^2} + \fft{\log(z)}{6(z-1)}\,,\nn\\
b_1&=& \fft{z-1}{4z} + \fft{\log z }{6z}\,,\nn\\
c_1&=& \fft{(z-1)^2}{12z^2\,(2z-1)} + \fft{\log z }{3(2z-1)}
\,,\nn\\
f_1&=&\fft{4-9z}{12z\, (2z-1)} + 
\fft{(1-2z+2z^2)\,\log z }{6z\,(z-1)\, (2z-1)}\,,
\eea
%%%%%%%
and
%%%%%
\bea
a_2&=&\fft{15 + 268z - 431 z^2 - 272 z^3 + 675 z^4}{4320z^4} +
\fft{7\psi(2, 1-z)}{60(z-1)}\nn\\
&&+\fft{(-20+26z+ 31z^2 -97z^3 + 60z^4 + z^3\,(21z-26)\,\log z )\,
\log z }{360z^3\, (z-1)^2}\,,\nn\\
b_2&=&\fft{16+38z - 147 z^2 + 989 z^3 - 2903 z^4 + 
1755 z^5}{4320z^4\, (z-1)} + \fft{7\psi(2,1-z)}{60z}\nn\\
&&+\fft{(35-127 z + 233 z^2 - 120 z^3 + (z-1)^2\, (21z-5)\, \log z )
\,\log z }{360z^2\, (z-1)^2}\,,\nn\\
c_2&=&\fft{-15 + 247 z - 1500 z^2 + 4316 z^3 - 7795 z^4 + 10359 z^5 -
8600 z^6 + 3240 z^7}{4320z^4\, (z-1)\, (2z-1)^2}\nn\\
&&-\fft{(-20 + 131 z - 306 z^2  + 274 z^3  + 24 z^4  - 202 z^5  
+ 120 z^6 )\, \log z }{360z^3\, (z-1)^2\, (2z-1)^2}\nn\\
&& +\fft{(42z-31)\,\log^2 z}{180(2z-1)} + 
\fft{7\psi(2,1-z)}{30(2z-1)}\nn\\
f_2&=& -\fft{4 + 12 z - 324 z^2  + 1216 z^3  + 209 z^4  - 4720 z^5 
+ 3240 z^6}{4320 z^4\, (2z-1)^2}\nn\\ 
&&+ \fft{7(1-2z+2z^2)\, \psi(2,1-z)}{60z\,(z-1)\,(2z-1)}
-\fft{(20 - 144 z + 351 z^2  - 301 z^3  + 60 z^4)\,\log z }{
180z^2\, (z-1)\, (2z-1)^2}\nn\\
&&+\fft{(-5 + 61 z - 165 z^2  + 250 z^3  - 230 z^4  + 84 z^5)\,
\log^2 z}{360z^2\, (z-1)^2\, (2z-1)^2}\,.
\eea
%%%%%%%
Here $z=\ft12(1+r^2)$ and $\psi(2, x)\equiv -\int^z\td z^{-1}\,
\log(1-\td z)\, d\td z$ is the di-logarithm function.

At large distance, we have
%%%%%
\bea
&&a=\fft{r}{\sqrt2}\Big(1 -\ft14\epsilon +\ft5{32}\epsilon^2\Big)
\,,\qquad
b=\fft{r}{\sqrt2}\Big(1 + \ft14\epsilon + \ft{13}{32}\epsilon^2\Big)
\nn\\
&&c=(1+\ft3{16}\epsilon^2)\, r\,,\qquad
f=\ft12(1-\ft3{16}\epsilon^2)\, r\,,\qquad
h=1 + \fft{3}{16}\, \epsilon^2\,.
\eea
%%%%%

            As a verification, one can check that the above cone
metric matches precisely to the conifolds obtained in the previous
section, up to $\epsilon^2$ order, expanding around $\phi=\pi$.

    At small distance, it can be matched with Case 2 in section
\ref{resolvsec}, provided that $q=-2/3 + \epsilon -
\ft{277}{270}\epsilon^2$.

   Again, we are seeing that an orderly perturbative expansion can  be
developed, with corrections to the metric functions at order $\ep$ and
$\ep^2$ that are regular both at short distances and at large
distances.  This again provides analytical evidence supporting the
findings from our numerical analysis, that regular AC solutions of the
first-order equations (\ref{fo2}) should exist for all $k$ and
$\ell$.

\section{More general 7-metrics of $G_2$ holonomy}\label{g2section}

   Having studied more general cohomogeneity 8-metrics with Spin(7) 
holonomy, we now turn to the consideration of analogous
generalisations for 7-metrics of $G_2$ holonomy.

\subsection{New $G_2$ metrics on $\R^3$ bundle over 
$\CP^2$}\label{g2cp2sec}

   We start from the left-invariant 1-forms $L_A{}^B$ of $SU(3)$, and
define complex 1-forms $\sigma\equiv L_1{}^3$, $\Sigma\equiv L_2{}^3$
and $\nu\equiv L_1{}^2$, as in section \ref{d8fosec}.  Defining
real 1-forms via $\sigma=\sigma_1 + \im\, \sigma_2$, etc, we then make
the ansatz
%%%%%
\be
ds_7^2 = dt^2 + a^2\, \sigma_i^2 + b^2\, \Sigma_i^2 + c^2\, \nu_i^2\,.
\label{new7met}
\ee
%%%%%
This is very like the ansatz for eight-dimensional Spin(7) metrics in 
(\ref{d8ansatz}),
except that the extra $U(1)$ direction $f^2\, \lambda^2$ in equation
(3.2) there is dropped.  We can therefore read off the results for
curvature, $T$ and $V$ from section 2 of \cite{cglphyper}, and
reproduced in (\ref{fo2}), by
dropping all the $f$ terms.  Thus we have
%%%%%
\bea
T &=& 2 {\a'}^2 + 2 {\beta'}^2 + 2{\gamma'}^2 + 8\a'\, \beta' + 2
\beta'\, \gamma' + 2 \a'\, \gamma'\nn\,,\\
V &=& - \fft{12}{a^2} - \fft{12}{b^2} - \fft{12}{c^2} +
\fft{2a^2}{b^2\, c^2} + \fft{2 b^2}{a^2\, c^2} + \fft{2 c^2}{a^2\,
b^2}\,.
\eea
%%%%%
Note that the principal orbits are the coset space $SU(3)/(U(1)\times
U(1))$, which is the six-dimensional flag manifold.

    We find that $V$ can be derived from the superpotential
%%%%%
\be
W = 4 a\, b\, c\, (a^2+b^2+c^2)\,.
\ee
%%%%%
From this, we arrive at the first-order equations
%%%%%
\be
\fft{\dot a}{a} = \fft{b^2+c^2-a^2}{a\, b\, c}\,,\quad
 \fft{\dot b}{b} = \fft{c^2+a^2-b^2}{a\, b\, c}\,,\quad
\fft{\dot c}{c} = \fft{a^2+b^2-c^2}{a\, b\, c}\,.\label{firstorder}
\ee
%%%%%

   It should be noted that these are identical to one of the sets of
first-order equations that can be derived for the triaxial Bianchi IX
system in $D=4$, with $ds_4^2 = dt^2 + a^2\, \sigma_1^2 + b^2\,
\sigma_2^2 + c^2\, \sigma_3^2$, where here the $\sigma_i$ are the
left-invariant 1-forms of $SU(2)$.  Specifically, they coincide with
the $D=4$ equations that correspond to the Nahm equations for the
``spinning top.''  This is the first-order system that admits
Eguchi-Hanson as a non-singular solution if $a=b$.  For unequal $a$,
$b$ and $c$, the system studied in \cite{begipapo}, and the general
solution was obtained. It was found that the associated Ricci-flat
metrics were singular when the three functions were unequal.

   We can use the same method here to solve the first-order equations
(\ref{firstorder}).  Thus we let $u=a\, b$, $v=b\, c$ and $w=c\, a$.
After defining a new radial coordinate $r$ by $dr=a\, b\, c\, dt=
\sqrt{u\, v\, w}\, dt$, we then get
%%%%%
\be
\fft{du}{dr} = \fft{2}{u}\,,\quad 
\fft{dv}{dr} = \fft{2}{v}\,,\quad 
\fft{dw}{dr} = \fft{2}{w}\,,
\ee
%%%%%
with the general solution
%%%%%
\be
u^2 = 4(r-r_1)\,,\quad v^2 = 4(r-r_2)\,,\quad w^2 =4(r-r_3)\,,
\ee
%%%%%
where $r_1$, $r_2$ and $r_3$ are constants of integration.\footnote{We
understand that the first-order equations (\ref{firstorder}) for the
Spin(7) metrics (\ref{new7met}) have also been obtained independently
by R. Cleyton (PhD thesis, Odense University), and by A. Dancer and
M.Y. Wang, who also noted that they are equivalent to the Nahm
equations, and hence are integrable.} The metric is
%%%%%
\be
ds_7^2 = \fft{dr^2}{u\, v\, w} + \fft{u\,w}{v}\, (\sigma_1^2 +
\sigma_2^2) + \fft{u\, v}{w}\, (\Sigma_1^2 + \Sigma_2^2) + \fft{v\,
w}{u}\, (\nu_1^2 + \nu_2^2)\,.
\ee
%%%%%
It can be seen that this is singular unless two of the $r_i$ are set
equal.  If two are set equal, so that $a=b$, we get, after a coordinate
transformation, the previously-known $G_2$ metric on the $\R^3$ bundle
over $\CP^2$.  

    The $G_2$ holonomy can be checked by looking for a
covariantly-constant spinor.  Equivalently, we can check to see if
there is a covariantly-constant 3-form (the calibrating form).  From
the exterior derivatives of the complex 1-forms given in the
hyper-K\"ahler paper, we can easily verify that 
%%%%%
\be
d(\sigma\wedge\bar\sigma) = -d(\Sigma\wedge\bar\Sigma) =
d(\nu\wedge\bar\nu) = -2\im\, \Re(\bar\sigma\wedge \Sigma\wedge\nu)\,.
\ee
%%%%%
From this we see that the 3-form $G_\3$, defined by
%%%%%
\be
G_\3 \equiv a\, b\, c\, \Re(\bar\sigma\wedge\Sigma\wedge\nu) + \im\, 
(-a^2\, \sigma\wedge\bar\sigma + b^2\, \Sigma\wedge\bar\Sigma + c^2\,
\nu\wedge \bar\nu)\label{g3}
\ee
%%%%%
is closed, $dG_\3=0$, by virtue of the first-order equations
(\ref{firstorder}).  A more complete calculation should show that it
is in fact covariantly constant.

   Note that the vielbein components of $G_\3$ will be constants.

\subsection{$G_2$ metric on $\R^3$ bundle over $S^4$ revisited}
\label{g2s4sec}

     We could attempt a similar more general construction of metrics
on the $\R^3$ bundle over $S^4$.  As we shall see, this does not in
fact seem to be possible.  It does, however, provide us with a more
convenient way of writing the standard $G_2$ metric on this manifold.

   Our starting point is the left-invariant 1-forms $L_{AB}$ on
$SO(5)$, introduced in section 2.  In the earlier discussion we
identified $P_a$ and $(R_1,R_2,R_3)$ as the 1-forms in the coset
$S^7=SO(5)/SU(2)_L$.  We now divide out by a further $U(1)$ factor,
associated with the 1-form $R_3$.  The
required $\CP^3$ principal orbits for the $\R^3$ bundle over $S^4$ are
thus described by the coset
%%%%%
\be
\CP^3 = \fft{SO(5)}{SU(2)_L\times U(1)_R}\,.
\ee
%%%%%

    From (\ref{so5d}), we can see that the following exterior
derivatives lie entirely within the coset:
%%%%%
\bea
d(P_0\wedge P_3 + P_1 \wedge P_2) &=& - 2R_1\wedge (P_0\wedge P_2 +
P_3\wedge P_1) + 2R_2\wedge (P_0\wedge P_1 + P_2\wedge P_3)\,,\nn\\
d(R_1\wedge R_2) &=&   \ft12 R_1\wedge (P_0\wedge P_2 +
P_3\wedge P_1) - \ft12 R_2\wedge (P_0\wedge P_1 + P_2\wedge P_3)\,.
\eea
%%%%%
In particular, we see that
%%%%%
\be
d(P_0\wedge P_3 + P_1\wedge P_2 + 4 R_1\wedge R_2) =0\,.
\ee
%%%%%
This corresponds to the nearly-K\"ahler structure on $\CP^3$.  Note,
however, that there is no result lying purely within the coset if we
try giving the $P_0\wedge P_3$ and $P_1\wedge P_2$ terms different
coefficients.  Thus we cannot break the $S^4$ base (whose coset
1-forms are $P_a$) apart.  This is quite different from the previous
example in section \ref{g2cp2sec}.

   The most general metric ansatz we can consider is therefore
%%%%%
\be
ds_7^2 = dt^2 + a^2\, (R_1^2 + R_2^2) + b^2\, P_a^2\,.
\ee
%%%%%
This is equivalent to the standard ansatz for the $G_2$ metrics on the
$\R^3$ bundle over $S^4$ \cite{brysal,gibpagpop}.

   The natural $SO(5)$-invariant ansatz for the calibrating 3-form is
%%%%%
\bea
G_\3 &=& dt\wedge [a^2 R_1\wedge R_2 - b^2 (P_0\wedge P_3 + P_1\wedge
P_2) ]\nn\\
&& + a\, b^2\, [R_2\wedge (P_0\wedge P_1 + P_2\wedge P_3) -
 R_1\wedge (P_0\wedge P_2 + P_3\wedge P_1)]\,.
\eea
%%%%%
From the condition $dG_\3=0$ we get
%%%%%
\be
\fft{d(a\, b^2)}{dt} = -\ft12 a^2 - 2 b^2\,,
\ee
%%%%%
while from $d{*G_\3}=0$ we get
%%%%%
\be
2b\, \fft{db}{dt}  + a=0\,,\qquad \fft{d(a^2\, b^2)}{dt} + 4 a\,
b^2=0\,.
\ee
%%%%
These imply the first-order equations $\dot a = \ft12 a^2\, b^{-2} -2$
and $\dot b = -\ft12 a\, b^{-1}$, which are the same, after
appropriate adjustment for normalisations, as those obtained
in \cite{cglphyper} for the $\R^3$ bundle over $S^4$.  The solution
can be written as 
%%%%%
\be
ds_7^2 = \Big(1-\fft{\ell^4}{r^4}\Big)^{-1}\, dr^2 + r^2\,
\Big(1-\fft{\ell^4}{r^4}\Big)\, (R_1^2 + R_2^2) + \ft12 r^2\, P_a^2\,.
\ee
%%%%%
 
\subsection{$G_2$ metrics for the six-function triaxial $S^3\times S^3$
ansatz}\label{g2sixfunsec}  

   Another class of $G_2$ metrics that may be studied has principal
orbits that are $S^3\times S^3$.  A rather general ansatz involving
six radial functions was considered in \cite{cglpg2,brgogugu}, and
first-order equations for $G_2$ holonomy were derived.
The metric for the six-function $G_2$ space is given by
%%%%%
\be
ds_7^2 = dt^2 + a_i^2\, (\sigma_i-\Sigma_i)^2 + b_i^2\, (\sigma_i +
\Sigma_i)^2\,,\label{7met2}
\ee
%%%%%
where $\sigma_i$ and $\Sigma_i$ are left-invariant 1-forms for two
$SU(2)$ group manifolds.  It was found that for $G_2$ holonomy,  
$a_i$ and $b_i$ must satisfy the first-order equations
%%%%%
\bea
\dot a_1 &=& \fft{a_1^2}{4 a_3\, b_2} + \fft{a_1^2}{4 a_2\, b_3}
      - \fft{a_2}{4b_3}  -\fft{a_3}{4b_2} - \fft{b_2}{4 a_3} -
        \fft{b_3}{4a_2}\,,\nn\\
\dot a_2 &=& \fft{a_2^2}{4 a_3\, b_1} + \fft{a_2^2}{4 a_1\, b_3}
      - \fft{a_1}{4b_3}  -\fft{a_3}{4b_1} - \fft{b_1}{4 a_3} -
        \fft{b_3}{4a_1}\,,\nn\\
%%%%%
\dot a_3 &=& \fft{a_3^2}{4 a_2\, b_1} + \fft{a_3^2}{4 a_1\, b_2}
      - \fft{a_1}{4b_2}  -\fft{a_2}{4b_1} - \fft{b_1}{4 a_2} -
        \fft{b_2}{4a_1}\,,\nn\\
\dot b_1 &=& \fft{b_1^2}{4 a_2\, a_3} - \fft{b_1^2}{4 b_2\, b_3}
      - \fft{a_2}{4a_3}  -\fft{a_3}{4a_2} + \fft{b_2}{4 b_3} +
        \fft{b_3}{4b_2}\,,\label{sixfo}\\
\dot b_2 &=& \fft{b_2^2}{4 a_3\, a_1} - \fft{b_2^2}{4 b_3\, b_1}
      - \fft{a_1}{4a_3}  -\fft{a_3}{4a_1} + \fft{b_1}{4 b_3} +
        \fft{b_3}{4b_1}\,,\nn\\
\dot b_3 &=& \fft{b_3^2}{4 a_1\, a_2} - \fft{b_3^2}{4 b_1\, b_2}
      - \fft{a_1}{4a_2}  - \fft{a_2}{4a_1} + \fft{b_1}{4 b_2} +
        \fft{b_2}{4b_1}\,,\nn
\eea
%%%%%

    One can look for solutions with regular Taylor expansions
corresponding to a collapsing $S^1$, $S^2$ or $S^3$ at $t=0$.  We find
no such regular solutions for a collapsing $S^1$ or $S^2$, but for a
collapsing $S^3$, we find that solutions which are regular near the
associated $S^3$ bolt at $t=0$ have a Taylor expansion with three free
parameters, and are given by
%%%%%%
\be
a_i = a_0 + \ft{1}{16a_0}\, t^2 + \cdots\,,\qquad
b_i = -\ft14 t + q_i\, t^3 + \cdots\,,
\ee
%%%%%%
where $a_0^{-2} = 64(q_1 + q_2 + q_3)$ (implying that
$q_1+q_2+q_3>0$).   A numerical analysis now shows that regularity at
large distance requires that
%%%%%%
\be
q_1\ge q_2=q_3\,,\qquad \hbox{or cyclic order}\,.
\ee
%%%%%%%%%
Thus the only regular solutions of the six-function equations
(\ref{sixfo}) are solutions also of the reduced 
four-function equations first obtained in \cite{brgogugu}.  Setting
$a_2=a_3$ and $b_2=b_3$ in (\ref{sixfo}), these are
%%%%%
\bea
\dot a_1&=& \fft{a_1^2}{2 a_2\, b_2} - \fft{a_2}{2b_2} - \fft{b_2}{2
a_2} \,,\nn\\
\dot a_2&=& \fft{a_2^2}{4 a_1\, b_2} - \fft{a_1}{4b_2} - \fft{b_2}{4
a_1} - \fft{b_1}{4 a_2}\,,\nn\\
\dot b_1&=& \fft{b_1^2}{4a_2^2} - \fft{b_1^2}{4 b_2^2}\,,\nn\\
\dot b_2&=& \fft{b_2^2}{4 a_2\, a_1} - \fft{a_2}{4a_1} - \fft{a_1}{4
a_2} + \fft{b_1}{4 b_2}\,.
\eea
%%%%%
Making the redefinitions
%%%%%
\be
A=\fft{a_2^2}{a_1^2}\,,\qquad
B=\fft{b_2^2}{a_1^2}\,,\qquad
F=\fft{b_1\, a_2\, b_2}{a_1^3}\,,\qquad
G=\fft{a_2\, b_2}{a_1}\,,
\ee
%%%%%%
the equations become
%%%%
\bea
A' &=&3A^2 + A(B-3) + F\,,\qquad
B'=3B^2 + B(A-3) - F\,,\nn\\
F'&=& (-4 + 3A + 3B)\, F \,,\quad
G^{-1}\, G'=-2 + A + B + \fft{F}{2A} - \fft{F}{2B}\,,
\eea
%%%%%
where a prime denotes a derivative with respect to $\eta$, which is
defined by $dt=2G\, d\eta$.  Note that $G$ is decoupled from the first
three equations.  If we now define 
%%%%%
\be
A=X+Y\,,\qquad  B=X-Y\,,
\ee
%%%%%
the first three equations give
%%%%%
\be
X'=4X^2-3X +2Y^2\,,\quad Y'=6X\, Y -3Y + F\,,\quad F'=2(3X-2)\, F\,.
\ee
%%%%%
By calculating $Y''$, using the other first-order equations, and then
writing $d/d\eta = X'\, d/dX$, we get the following second-order
equation:
%%%%%
\bea
&&(4X^2-3X+2Y^2)\, \Big[ (4X^2-3X+2Y^2)\, \fft{d^2Y}{dX^2} + 4Y\,
\Big(\fft{dY}{dX}\Big)^2 -4(X-1)\, \fft{dY}{dX}\Big] \nn\\
&&+12(X+Y-1)\, (X-Y-1)\, Y =0\,.
\eea
%%%%%
Note that a special solution of this equation is
%%%%%
\be
Y=\fft{\sqrt{3-4X}}{2}\,.
\ee
%%%%%
This is in fact the isolated
solution that was found in \cite{brgogugu}.  It can be written as
%%%%%
\bea
&&a_1 = -\ft12 r\,,\qquad
a_2 = \ft14\sqrt{3(r-\ell)(r+3\ell)}\,,\nn\\
&& b_1 = \ell\,
   \fft{\sqrt{r^2-9\ell^2}}{\sqrt{r^2-\ell^2}}\,,\qquad
b_2 = -\ft14\, \sqrt{3(r+\ell)(r-3\ell)}\,,\label{simsol}
\eea
%%%%%
where $dt=-\ft32\ell\, dr/b_1$. 

   Taking the $q_2=q_3$, our numerical analysis shows that $q_2/q_1\ge
1$ is a non-trivial parameter characterising inequivalent solutions,
which are non-singular and ALC when
%%%%%
\be
-\ft12 < \fft{q_2}{q_1} <1\,.\label{g2fam}
\ee
%%%%%
The limiting case $q_2/q_1=1$ gives an AC solution, which
is in fact the previously-known $G_2$ metric on the spin bundle of
$S^3$ \cite{brysal,gibpagpop}.  The general family in (\ref{g2fam})  
includes the specific explicitly-known example (\ref{simsol}) found in
\cite{brgogugu}.  Converting to the proper-distance coordinate $t$, we
find that the solution (\ref{simsol}) corresponds to
$q_2/q_1=-\ft1{14}$.  

   Our numerical analysis supports the perturbative arguments given in
\cite{brgogugu}, which indicated the existence of the non-trivial
1-parameter family of ALC solutions that we have found numerically.
By analogy with our notation for the new ALC 8-manifolds of Spin(7)
holonomy found in \cite{cglpspin7}, we shall denote the explicit $G_2$
solution (\ref{simsol}) of \cite{brgogugu} by $\bB_7$.  We shall also 
denote the 1-parameter family of non-singular ALC solutions with $-\ft12
< q_2/q_1<-\ft1{14}$ by $\bB_7^-$, and those with $-\ft1{14} < q_2/q_1
<1$ by $\bB_7^+$.  It should be noted, however, that there is no
$\bA_7$ solution of $\R^7$ topology, which would be analogous to the
$\bA_8$ solution on $\R^8$ found in \cite{cglpspin7}.  This is because
unlike the metrics studied in \cite{cglpspin7}, where the principal
orbits were spheres ($S^7$), which have the possibility of collapsing
down smoothly to a point at the original of spherical polar
coordinates, here the principal orbits are $S^3\times S^3$, and so a
smooth collapse to a point is impossible.

   Note that besides the upper bound $q_2/q_1=1$ when we recover the
original AC metric of $G_2$ holonomy \cite{brysal,gibpagpop} on the
$\R^4$ bundle over $S^3$, the lower bound, $q_2/q_1=-\ft12$,
corresponds to the Gromov-Hausdorff limit in which we get $M_6\times
S^1$, where $M_6$ is the Ricci-flat K\"ahler metric on the deformed
conifold.  The various non-singular solutions are depicted in 
Figure 4 below.

\begin{figure}[ht]
\leavevmode\centering
\epsfxsize=5in
\epsfbox{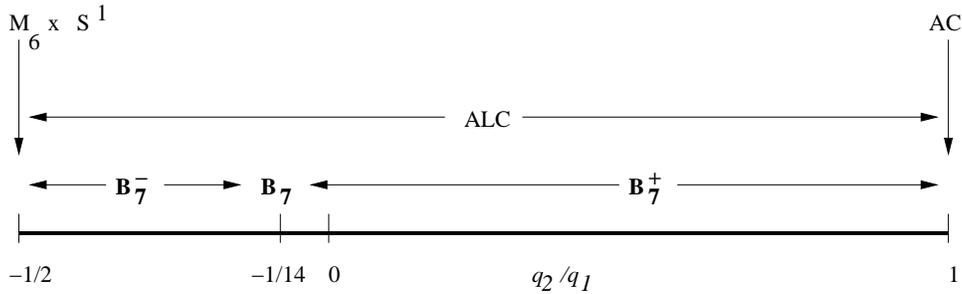}
\caption{The non-singular $G_2$ metrics $\bB_7$ and $\bB_7^\pm$ 
as a function of $q_2/q_1$}
\end{figure} 

\section{Conclusions}

    In this paper, we have made a rather extensive investigation of
many of the possible classes of metrics of cohomogeneity one in
dimensions eight and seven that might give rise to the exceptional
holonomies Spin(7) and $G_2$ respectively.  For the case of eight
dimensions, we considered first the situation where the principal
orbits are topologically $S^7$, endowed with a homogeneous metric
given by the coset $SO(5)/SO(3)$.  One can view such metrics as $S^3$
bundles over $S^4$, where the $S^3$ fibres are themselves required to
be only left-invariant under the action of $SU(2)$.  The
eight-dimensional metric ansatz therefore has an $SO(5)$ isometry, and
involves four functions of the radial variable; three characterising
homogeneous ``squashings'' of the $S^3$ fibres, and a fourth measuring
the radius of the $S^4$ base.  We obtained first-order equations for
these functions, coming from the requirement of Spin(7) holonomy, and
we then examined the possible solutions.  We found by a numerical
analysis that there should exist a family of complete and non-singular
metrics with a non-trivial parameter $\lambda^2\le 4$, 
which we denote by $\bC_8$, which are topologically $\R^2$ bundles
over $\CP^3$.  The parameter $\lambda$ characterises the degree of
squashing of the minimal $\CP^3$ bolt, with $\lambda^2=4$
corresponding to the Fubini-Study metric on $\CP^3$.  This limiting
case has $SU(4)$ holonomy, and the metric has been known for a long
time, but the metrics with $\lambda^2<4$ are new.  They are ALC, and
on the $S^3$ fibres they exhibit a similar behaviour to that seen in
the Atiyah-Hitchin metric in $D=4$.

     We then considered eight-dimensional metrics of cohomogeneity one
whose principal orbits are the Aloff-Wallach spaces
$N(k,\ell)=SU(3)/U(1)_{k,\ell}$.  We began with a more complete and
explicit discussion of the Einstein metrics on $N(k,\ell)$ than has
previously appeared in the literature.  Earlier results showed the
existence of Einstein metrics \cite{myw}, gave an explicit result for an
Einstein metric on $N(k,\ell)$ \cite{casrom}, and gave a
demonstration, based on the results of \cite{casrom}, that there exist
two inequivalent Einstein metrics on each $N(k,\ell)$ except $N(1,-1)$
\cite{pagpop}.  In this paper, we gave an explicit construction, from
first principles, of the two Einstein metrics, deriving them from the
conditions for weak $G_2$ holonomy.  These Einstein metrics form the
possible bases for cones in eight-dimensional AC metrics of Spin(7)
holonomy.  Although we were unable to obtain the general solutions of
the first-order equations for Spin(7) holonomy, we were able to find
an isolated  ALC solution explicitly for all $(k,\ell)$.  In general,
the metric will not be completely non-singular, but rather, will have
an orbifold structure, of the local form $\R^4/Z_p\times \CP^2$, where
$p=|k+\ell|$.  

    We also studied the solutions of the first-order equations for
$N(k,\ell)$ principal orbits numerically, and in certain perturbative
expansions, and found evidence for the existence of complete and
non-singular metrics, both AC and ALC, for all $(k,\ell)$.  

   We then turned our attention to seven-metrics of cohomogeneity one
with $G_2$ holonomy.  We studied the first-order system for the case
where the principal orbits are the flag manifold $SU(3)/(U(1)\times
U(1))$.  This can have three metric functions depending on the radial
variable.  We showed that the first-order equations implying $G_2$
holonomy reduce to the same ones that are encountered in one of the
first-order systems for hyper-K\"ahler Bianchi IX metrics in $D=4$,
and hence they can be solved by the same method that was used in
\cite{begipapo}.  As in that case, it turns out that the resulting
metrics are singular unless two of the metric functions are equal, in
which case the system reduces to the already-studied one whose
solution is the complete non-singular $G_2$ metric on the $\R^3$
bundle of self-dual 2-forms over $\CP^2$ \cite{brysal,gibpagpop}.

    A second $G_2$ example arises if the principal orbits are $\CP^3$,
described as an $S^2$ bundle over $S^4$.  Only two metric functions are
possible in this case, describing the radii of the $S^2$ fibres and
the $S^4$ base, and the system reduces to the one that was solved in
\cite{brysal,gibpagpop}, giving the non-singular metric on the $\R^3$
bundle of self-dual 2-forms over $S^4$.

    A third possibility is when the principal orbits are $S^3\times
S^3$, described as an $S^3$ bundle over $S^3$.  In principle one can
now write an ansatz with nine functions of the radial coordinate
\cite{cglpg2}, although it is not clear that a first-order system of
equations for $G_2$ holonomy can arise in this case.  A simpler system
with six functions (three measuring the radii of the squashed $S^3$
base, and three measuring the radii of the squashed $S^3$ fibres) was
also considered in \cite{cglpg2}, and in \cite{brgogugu}, for which a
first-order system implying $G_2$ holonomy exists.  Our numerical
investigations in this paper lead to the conclusion that the solutions
will only be non-singular if pairs of metric functions on the base and
fibre 3-spheres are set equal.  This results in a four-function
system, whose general solution has not been found analytically.  An
isolated ALC example was found in \cite{brgogugu}, and arguments for
the existence of a non-trivial 1-parameter family were presented.  We
have analysed the system numerically in this paper, and we also find
evidence for the existence of such a family of non-singular solutions.

\section*{Acknowledgements} 

    We are grateful to Michael Atiyah, Andrew Dancer, Krystoff
Galicki, Nigel Hitchin, Simon Salamon, James Sparks, Paul Tod and
McKenzie Wang for helpful discussions, and to K. Kanno and Y. Yasui
for pointing out an error in the original equations (\ref{weqs2}).
Subsets of the authors thank DAMTP (C.N.P.), the Benasque Centre for
Science (M.C., G.W.G., C.N.P.)  and the Ecole Normale (M.C., C.N.P.)
for hospitality at various stages during this work.  M.C. is supported
in part by DOE grant DE-FG02-95ER40893 and NATO grant 976951; H.L.~is
supported in full by DOE grant DE-FG02-95ER40899; C.N.P.~is supported
in part by DOE DE-FG03-95ER40917.

\section*{Note Added}

    After this paper was completed, the paper \cite{kanyas} appeared,
which also studies the solutions of the first-order equations in
\cite{cglphyper} for metrics of Spin(7) holonomy with $N(k,\ell)$
principal orbits.  This has an extensive overlap with our results in
sections 3.2 and 3.3.  In particular, \cite{kanyas} also obtains the
explicit ALC solutions (\ref{genmetrics}), and discusses the existence
of more general classes of ALC metrics.

\appendix

\section{The geometry of $SU(3)$}

    In order for the better understanding of the global structures
of the Aloff-Wallach
spaces $N(k,\ell)=SU(3)/U(1)_{k,\ell}$, it can be useful to have
available an explicit parameterisation of the group $SU(3)$.  One may
parameterise any $SU(3)$ group element $g$ in terms of generalised
Euler angles as
%%%%%
\be
g=U\, e^{\im\, \lambda_5\, \xi}\, \wtd U\,  
e^{\ft{\im\, \sqrt3}{2}\,\lambda_8\, \tau }\,,
\ee
%%%%%%
where
%%%%%
\be
U \equiv e^{\ft{\im}{2}\,\lambda_3\, \phi}\, 
        e^{\ft{\im}{2}\,\lambda_2\, \theta }\,
    e^{\ft{\im}{2}\, \lambda_3\, \psi }\,,\qquad
\wtd U \equiv 
    e^{\ft{\im}{2}\, \lambda_3\, \td\phi}\, 
    e^{\ft{\im}{2}\, \lambda_2\, \td\theta}\, 
    e^{\ft{\im}{2}\, \lambda_3\,\wtd \psi}\,,
\ee
%%%%%
where $(\theta,\phi,\psi)$ are Euler angles for $SU(2)$,
$(\td\theta,\td\phi,\wtd\psi)$ are Euler angles for another 
$SU(2)$, and the $\lambda_a$ are the Gell-Mann matrices.
The coordinate ranges are
%%%%%
\be
0\le \theta\le \pi\,,\qquad 0\le\td\theta\le \pi\,,\qquad
0\le\xi\le \ft12 \pi\nn
\ee
%%%%%
for the ``latitudes,'' while the azimuthal coordinates have the periods
%%%%%
\be\Delta\, \phi=2\pi\,,\quad \Delta\, \psi=4\pi\,,\quad
\Delta\, \td\phi=2\pi\,,\quad \Delta\, \wtd\psi=4\pi\,,\quad
\Delta\, \tau=2\pi\,.\label{azimuthal}
\ee
%%%%%
Since the determination of these periods is slightly non-trivial, and
errors have occurred in various published papers, we shall give an
explicit derivation of the periods below.

    It is useful to define left-invariant 1-forms $s_i$ and $\td s_i$ 
for the two $SU(2)$ subgroups in the standard way:
%%%%%
\be
U^{-1}\, dU =\ft{\im}{2}\, \lambda_i \, s_i\,,\qquad 
\wtd U^{-1}\, d\wtd U =\ft{\im}{2}\, \lambda_i \, \td s_i\,,
\ee
%%%%%
giving
%%%%%
\be
s_1=\cos\psi\, d\theta + \sin\psi\, \sin\theta\, d\phi\,,\quad
s_2=-\sin\psi\, d\theta + \cos\psi\, \sin\theta\, d\phi\,,\quad
s_3=d\psi+\cos\theta\, d\phi\,,
\ee
%%%%%
and similarly for $\td s_i$.
Calculating the $SU(3)$ left-invariant 1-forms $X_a$ defined by
$g^{-1}\, dg= \ft{\im}2 \lambda_a \, X_a$, and taking $\nu_1+\im\,
\nu_2 \equiv X_1+\im\, X_2$, $\sigma_1+\im\, \sigma_2 \equiv X_4+\im\,
X_5$, $\Sigma_1+\im\, \Sigma_2 \equiv X_6+\im\, X_7$, together with 
$\sigma_3\equiv X_3$ and $\sigma_8\equiv X_8$ for the two Cartan subalgebra
1-forms, we find
%%%%%
\bea
\nu_1+\im\, \nu_2 &=& \im\, \td s_1 +\td s_2 +
e^{\im\,\wtd\psi}\, \Big[ \im\, \cos\xi\, 
[(\cos\td\phi + \im\, \sin\td\phi\, \cos\td\theta)\, s_1 \nn\\
&&\qquad\qquad\qquad
 +(\sin\td\phi - \im\, \cos\td\phi\, \cos\td\theta)\, s_2] +
\ft14(3+\cos2\xi)\, \sin\td\theta\, s_3\Big] \,,\nn\\
\sigma_1 + \im\, \sigma_2 &=& \im\, e^{\fft{\im}{2} \wtd\psi 
+ \fft{3\im}{2}\tau}\, \Big[ 2e^{\fft{\im}{2}\, \td\phi}\, 
\cos\ft12\td\theta\, (d\xi - \ft{\im}{4}\, \sin2\xi\, s_3) 
+e^{-\fft{\im}{2}\, \td\phi}\, \sin\xi\, \sin\ft12\td\theta\, (s_1+\im\,
s_2)\Big]\,,\nn\\
\Sigma_1 + \im\, \Sigma_2 &=& \im\, e^{-\fft{\im}{2} \wtd\psi 
+ \fft{3\im}{2} \tau}\, \Big[ 2e^{\fft{\im}{2}\, \td\phi}\, 
\sin\ft12\td\theta\, (d\xi - \ft{\im}{4}\, \sin2\xi\, s_3) 
-e^{-\fft{\im}{2}\, \td\phi}\, \sin\xi\, \cos\ft12\td\theta\, (s_1+\im\,
s_2)\Big]\,,\nn\\
\sigma_3 &=& \td s_3 + \cos\xi\, \sin\td\theta\, (\sin\td\phi\, s_1 -
\cos\td\phi\, s_2) + \ft14(3+\cos2\xi)\, \cos\td\theta\, s_3\,,\nn\\
\sigma_8 &=& \sqrt3\, (d\tau -\ft12 \sin^2\xi\, s_3)\,.
\eea 
%%%%%

    For some purposes it is highly advantageous to introduce instead
right-invariant 1-forms $\td t_i$ for the second $SU(2)$ group, 
defined by $d\wtd U\,\wtd U^{-1}= \ft{\im}{2}\, \lambda_i\, \td t_i$. 
In terms of the Euler angles $(\td\theta, \td\phi,\td\psi)$, these are
given by
%%%%%
\be
\td t_1=\cos\td\phi\, d\td\theta 
              + \sin\td\phi\, \sin\td\theta\, d\wtd\psi\,,\quad
\td t_2=\sin\td\phi\, d\td\theta - \cos\td\phi\, 
         \sin\td\theta\, d\wtd\psi\,,\quad
\td t_3=d\td\phi+\cos\td\theta\, d\wtd\psi\,.
\ee
%%%%%
The $SU(3)$ left-invariant 1-forms $\nu_1$, $\nu_2$ and $\sigma_3$
then become
%%%%%
\bea
\nu_1+\im\, \nu_2 &=& e^{\im\, \wtd\psi}\, 
\Big[ \im\, (\cos\td\phi + \im\, \sin\td\phi\, \cos\td\theta)\,
[\td t_1 +\cos\xi\,  s_1]\\
&&\qquad 
+\im\, (\sin\td\phi - \im\, \cos\td\phi\, \cos\td\theta)\, [\td t_2 
+\cos\xi\, s_2] +\sin\td\theta\, 
[\td t_3 +\ft14(3+\cos2\xi)\,s_3] \Big] \,,\nn\\
\sigma_3 &=& \sin\td\theta\,\Big[  \sin\td\phi\, [\td t_1 +
\cos\xi\, s_1] -
\cos\td\phi\, [\td t_2 +\cos\xi \,s_2]\Big] 
+ \cos\td\theta\, [\td t_3 +\ft14(3+\cos2\xi)\, s_3]\,.\nn
\eea
%%%%%%

   From these results, it is straightforward to show that
%%%%%
\bea
ds_4^2\equiv 
\sigma_1^2 + \sigma_2^2 + \Sigma_1^2 + \Sigma_2^2 &=& 4(d\xi^2 +\ft14
\sin^2\xi\, (s_1^2 + s_2^2) + \ft14 \sin^2\xi\, \cos^2\xi\, s_3^2)\,,
\\
ds_3^2\equiv \nu_1^2+ \nu_2^2 + \sigma_3^2 &=& (\td t_1+\cos\xi\, s_1)^2 +
     (\td t_2+\cos\xi\, s_2)^2 + [\td t_3 +\ft14(3+\cos2\xi)\,
     s_3]^2\,.\nn
\eea
%%%%%
The metric $ds_4^2$ is 4 times the standard Fubini-Study metric on
$\CP^2$, and since its principal orbits at fixed $\xi$ are $SU(2)$,
this proves that $\phi$ and $\psi$ must indeed have the periods given
in (\ref{azimuthal}).  The 8-metric
%%%%%
\bea
ds_8^2 &=&\equiv \nu_1^2+ \nu_2^2 + \sigma_3^2 +\sigma_8^2 
   +\sigma_1^2 + \sigma_2^2 + \Sigma_1^2 + \Sigma_2^2 \nn\\
&=& (\td t_1+\cos\xi\, s_1)^2 +
     (\td t_2+\cos\xi\, s_2)^2 + [\td t_3 +\ft14(3+\cos2\xi)\,
     s_3]^2  \nn\\
&& +3(d\tau- \ft12\sin^2\xi\, s_3)^2   +4(d\xi^2 +\ft14
\sin^2\xi\, (s_1^2 + s_2^2) + \ft14 \sin^2\xi\, \cos^2\xi\, s_3^2)
\label{su3metric}
\eea
%%%%%
is then the canonical bi-invariant metric on $SU(3)$, viewed as a $U(2)$ 
bundle over $\CP^2$.  

   If we project the metric (\ref{su3metric}) orthogonally to
$\del/\del\tau$, which amounts to dropping the term $\sigma_8^2$, we
get a metric on the the Aloff-Wallach space $N(1,1)$, viewed as an
$SO(3)$ bundle over $\CP^2$ (see \cite{pagpop}).  The fact that the
bundle is $SO(3)$ and not $SU(2)$ means that $\td\phi$ and $\wtd\psi$
must indeed have the periods given in (\ref{azimuthal}).  (We see from 
(\ref{su3metric}) that we have an $SO(3)$ bundle as opposed to $SU(2)$
since $\td\phi$ has period $2\pi$.)
  
\section{The Atiyah-Hitchin system and the Ricci flow 
        on $SU(2)$}\label{ahsumsec}

    The Ricci flow equations, which are encountered when studying the
renomalisation group equations for the target-space metric $g_{ij}$ in a 
sigma model, are
%%%%%
\be
\fft{dg_{ij}}{d\mu} = R_{ij}\,,
\ee
%%%%%
where $R_{ij}$ is the Ricci tensor of $g_{ij}$.  For metrics of the form
%%%%%
\be
ds^2 = A\, \sigma_1^2 + B\, \sigma_2^2 + C\, \sigma_3^2\,,
\ee
%%%%%
the non-vanishing components of the Ricci tensor in the
triad $(\sigma_1,\sigma_2,\sigma_3)$ are
%%%%%
\be
R_{11} = \ft12 A\, (A^2 -(B-C)^2)\,,\quad
R_{22} = \ft12 B\, (B^2 -(C-A)^2)\,,\quad
R_{33} = \ft12 C\, (C^2 -(A-B)^2)\,.
\ee
%%%%%
The Ricci-flow equations are therefore
%%%%%
\be
\fft{2}{A}\, \fft{dA}{d\mu} = A^2-B^2 -C^2 + 2B\, C\,,
\ee
%%%%%
and cyclic permutations.

  If we drop the terms involving $b$ in (\ref{spin7fo}), we get the
Atiyah-Hitchin system
%%%%%
\bea
\dot a_1 &=& \fft{a_1^2 - (a_2-a_3)^2}{a_2\, a_3} \,,\nn\\
\dot a_2 &=& \fft{a_2^2 - (a_3-a_1)^2}{a_3\, a_1} \,,\nn\\
\dot a_3 &=& \fft{a_3^2 - (a_1-a_2)^2}{a_1\, a_2} \,.\label{atihit}
\eea
%%%%%
It is a curious fact that the Ricci-flow and the Atiyah-Hitchin system
go into themselves under the identification $A=a_1$, $B=a_2$ and
$C=a_3$, together with a suitable change of parameterisation, 
$dt= 2 a_1\, a_2\, a_3\, d\mu$.  As far as we are aware, this
coincidence between first-order equations coming from a superpotential
and the Ricci-flow equations occurs only in this case.  It follows
from the work of Atiyah and Hitchin that the Ricci-flow is completely
integrable in this case, and in what follows we shall review the
standard way of solving (\ref{atihit}).

    On begins by defining a new radial
coordinate $\eta$ by $dt=a_1\, a_2\, a_3\, d\eta$, and also
introducing new variables $w_i$ as in (\ref{wdef}).  One then has 
%%%%%
\be
\fft{d(w_1+w_2)}{d\eta} = 4 w_1\, w_2\,,\qquad
\fft{d(w_2+w_3)}{d\eta} = 4 w_2\, w_3\,,\qquad
\fft{d(w_3+w_1)}{d\eta} = 4 w_3\, w_1\,.\label{weqs}
\ee
%%%%%

   It has been observed that there is an $SL(2,\R)$ symmetry of this
system.   Namely, letting $a$, $b$, $c$ and $d$ be constants (nothing
to do with the previous metric functions!), then if we define
transformed variables $v_i$ in place of $w_i$, and a transformed
radial coordinate $\xi$ in place of $\eta$, according to
%%%%%
\be
\xi = \fft{a\, \eta + b}{c\, \eta + d}\,,\qquad
w_i = -\fft{c}{2(c\, \eta + d)} + \fft{1}{(c\, \eta+d)^2}\,
v_i\,,
\ee
%%%%%
where $a\, d -b\, c=1$, then the equations (\ref{weqs})
become
%%%%%
\be
\fft{d(v_1+v_2)}{d\xi} = 4 v_1\, v_2\,,\qquad
\fft{d(v_2+v_3)}{d\xi} = 4 v_2\, v_3\,,\qquad
\fft{d(v_3+v_1)}{d\xi} = 4 v_3\, v_1\,.\label{veqs}
\ee
%%%%%
This allows one to transform a given solution into another, using the
$SL(2\R)$. 

   The equations (\ref{weqs}) that give the Atiyah-Hitchin metric can
be solved by defining a new radial coordinate $r$, related to $\eta$
by $dr= u^2\, d\eta$, with $u$ being a solution of
%%%%%
\be
\fft{d^2u}{dr^2} + \ft14 u\, \cosec^2 r  = 0\,.\label{elleq}
\ee
%%%%%
It can then be verified that the solution is
%%%%%
\bea
w_1 &=& - u\, u' - \ft12 u^2\, \cosec r\,,\nn\\
w_2 &=&  -u\, u' + \ft12 u^2\, \cot r\,,\\
w_3 &=&  - u\, u' + \ft12 u^2\, \cosec r\,,\nn
\eea
%%%%%
where $u'$ means $du/dr$.  The correct solution of (\ref{elleq}) to
choose for $u$ is
%%%%%
\be
u= \sqrt{2\sin r}\, K(\sin\ft12 r)\,,
\ee
%%%%%
where
%%%%%
\be
K(k) \equiv \int_0^{\pi/2} \fft{d\phi}{(1-k^2\, \sin^2\phi)^{1/2}}
\ee
%%%%%

\section{Proof of positivity of $Q(A,B)$}\label{qsec}

    In section \ref{2einsec}, our proof that the conditions for
Einstein metrics on $N(k,\ell)$ implied precisely the same set of
solutions as the ostensibly more restrictive conditions for weak $G_2$
holonomy depended upon the assertion that the function $Q(A,B)$ in
(\ref{ppoly}) is non-vanishing for all real positive $A$ and $B$.
We present a proof of this property here.

   The sixth-order polynomial $Q(A,B)$ occurring in section
\ref{2einsec} is given by
%%%%%
\bea
Q(A,B)&=& 5A^6 -6A^5 -5A^4 +12A^3 -5A^2-6A \nn\\
&&+5B^6 -6B^5 -5B^4 +12B^3 -5B^2-6B \nn\\
&&-A\, B\, (6A^4 -42 A^3 +36 A^2 +36 A + 6B^4 -42 B^3 +36B^2 +36B 
-42)\nn\\
&& -A^2\, B^2\, (5A^2+36A+5B^2+36B -12A\, B -130) +5\,.\label{Qdef}
\eea
%%%%%
In order to show that all solutions of the Einstein equations for the
seven-dimensional $N(k,\ell)$ spaces are also solutions of the weak
$G_2$ holonomy equations, we need to establish that $Q(A,B)$ is
non-vanishing whenever $A$ and $B$ are both positive.  To do this, we 
define $A=x+y$, $B=x-y$, in terms of which (\ref{Qdef}) becomes
%%%%%
\be
Q=(4x^2+1)(12x^2-12x+5) + 4(16x^4+48x^3-80x^2+36x-13)\, y^2 +
64(4x^2-6x+3)\, y^4\,.\label{Qdef2}
\ee
%%%%%
Note that since $A$ and $B$ are positive, it follows that $x>0$,
and although $y$ can have either sign, it appears only via $y^2$ and $y^4$.
Solving $Q=0$ for $y^2$, we get
%%%%%
\be
y^2 = \fft{13-36x+80x^2-48x^3-16x^4\pm(1-4x^2)\, \sqrt{J}}{
             32(3-6x+4x^2)}\,,
\ee
%%%%%
where $J\equiv 16x^4 + 96x^3 -200x^2 +120x -71$.  For $y^2$ to be real
we must therefore have $J\ge0$ or $x=\ft12$.  To have $J\ge0$ (and
$x$ positive) we must have $x\ge 1.2873\ldots$.  Now the coefficients
of $y^0$ and $y^4$ in (\ref{Qdef2}) are positive for all real $x$, and
the coefficient of $y^2$ is positive for all $x> .898374\ldots$.  Thus
$Q$ is positive for all $x$ that satisfies $J\ge0$.  The case
$x=\ft12$ implies $y^2=\ft14$ and hence $(A,B)=(1,0)$ or $(0,1)$, both
of which violate the requirement of $A$ and $B$ both being positive.
Thus we have proved that $Q$ is positive (and hence non-vanishing)
whenever $A$ and $B$ are both positive.   This completes the proof.

\end{document}